\documentclass[twocolumn]{rbef}
\usepackage{lipsum}
\usepackage{bbm}
\usepackage{subfig}

\numeracao{xxxx}
\volume{xx}
\numero{xx}
\ano{2019xx}
\doi{http://dx.doi.org/xxxx}
\tipodeartigo{Artigos Gerais}

	\author[*]{J. A. S. Lima}
	\author[**]{R. C. Santos}
\affil[*]{Universidade de São Paulo, Departamento de Astronomia (IAG-USP)\thanks{\href{emailto:jas.lima@iag.usp.br}{jas.lima@iag.usp.br}}}
\affil[**]{Universidade Federal de São Paulo, Campus Diadema, Departamento de Física\thanks{\href{emailto:rose.clivia@unifesp.br}{rose.clivia@unifesp.br}}}
	
\titulo{Do Eclipse Solar de 1919 ao Espetáculo das Lentes Gravitacionais}

\begin{document}

\begin{primeirapagina}

\begin{abstract}
\noindent Um século depois de observada a deflexão da luz emitida pelas estrelas distantes, durante o eclipse solar de 1919, é interessante saber os conceitos que emergiram do experimento e suas consequências teóricas e observacionais para a cosmologia e astrofísica moderna. Além de confirmar a teoria gravitacional de Einstein, o maior legado foi a construção de uma nova área de pesquisa na ciência do cosmos,  conhecida como lentes gravitacionais. A formação e magnificação de imagens múltiplas (miragens) pelo campo gravitacional de uma lente compacta ou extensa, estão entre os mais impressionantes fenômenos da natureza. Neste artigo apresentamos uma visão pedagógica do primeiro efeito genuíno de lente gravitacional, o quasar duplo QSO 0957+561. Descrevemos  a formação de anéis,  dos arcos gigantes, pequenos arcos e imagens múltiplas de Supernovas. É também surpreendente que a constante de Hubble e a quantidade de matéria escura no Universo possam ser mensuradas pela mesma técnica. Finalmente, o lenteamento de ondas gravitacionais, um efeito possível mas ainda não detectado, será também brevemente discutido. 

\end{abstract}

	\begin{otherlanguage}{english}
\begin{abstract}
\begin{center}
\textbf{\Large From Solar Eclipse of 1919 to the Spectacle of Gravitational Lensing}
\end{center}
\vspace{0.5cm}
\noindent  A century after observing the deflection of light emitted by distant stars during the solar eclipse of 1919, it is interesting to know the concepts emerged from the experiment and the theoretical and observational consequences for modern cosmology and astrophysics. In addition to confirming Einstein's gravitational theory, its greatest legacy was the construction of a new research area to cosmos science dubbed gravitational lensing. The  formation and magnification of multiple images (mirages) by the gravitational field of a compact or extended lens are among the most striking phenomena of nature. This article presents a pedagogical view of the first genuine gravitational lens effect, the double quasar QSO 0957 + 561. We also describe the formation of rings, giant arcs, arclets and multiple Supernova images. It is also surprising that the Hubble constant and the amount of dark matter in the Universe can be measured by the same technique. Finally, the lensing of gravitational waves, a possible but still not yet detected effect, is also briefly discussed.
\end{abstract}

	\end{otherlanguage}

\end{primeirapagina}
\saythanks

\section{Introdução}

O efeito de lente gravitacional (LG) refere-se a deflexão da luz pelo campo gravitacional da distribuição de matéria no Universo. Sua descrição pressupõe a existência do observador e de uma fonte  luminosa distante, intercalados por uma lente gravitacional. A lente defletora pode ser  um único astro ou um conjunto deles (planetas, estrelas, galáxias ou aglomerado de galáxias).  

O fenômeno é acromático, ou seja, não depende do comprimento de onda. Além disso,  difere das lentes na ótica convencional porque o foco não é bem definido. As imagens múltiplas, ou talvez melhor, as miragens gravitacionais,  apresentam distorções, amplificações e atrasos relativos (\textit{time delay})  na propagação da luz até as imagens. Essas diferenças de tempo são mensuráveis quando um processo transiente ocorre na fonte.

Os primeiros passos da nova área foram impulsionados pela deflexão da luz observada no eclipse solar total de 1919 \cite{Apais,Crispino2019}. Contudo, foi somente a partir dos anos 60-70, que o desenvolvimento teórico e possíveis aplicações do efeito de LGs progrediram mais rapidamente. 

Em 1979  (60 anos após o eclipse de 1919!), foi observada a primeira formação de uma imagem dupla de uma fonte distante. Tal resultado, teve grande impacto nas aplicações de interesse astrofísico e cosmológico \cite{ELTurner}. Eventos observados nas últimas décadas demonstraram inequivocamente que as LGs estão entre os fenômenos mais espetaculares da natureza. 

A teoria de lentes gravitacionais\footnote{O nome LG decorre da semelhança do fenômeno com a propagação da luz em meios materiais refratores (vidro, água, etc.). Na aproximação pós-newtoniana da relatividade geral, mostra-se que a velocidade aparente da luz, na presença  da gravitação, é menor do que seu valor no vácuo (c'=c/n), ou seja, o índice de refração efetivo $n$ é maior do que a unidade (ver \textbf{Apêndice A}).} tem sido discutida com diferentes enfoques em muitos livros textos \cite{SEF1999,Mol2002,CK2018} e artigos de revisão \cite{NB1996,Schneider2006,B2010,T2018} (antigos e recentes).   Genericamente, as LGs são  separadas em 2 classes distintas: (i) \textit{macrolentes} e (ii) \textit{microlentes}.  

As macrolentes estão subdivididas em 2 tipos: Fortes  e Fracas. As lentes fortes são observadas nos aglomerados de galáxias e aparecem como arcos gigantes (\textit{giant arcs}) e pequenos  arcos (\textit{arclets}). Em geral, provocam imagens múltiplas, o que implica na presença de uma ou mais lentes com intenso campo gravitacional.  O caso de lentes fracas ocorre quando o campo não é forte o suficiente, tal como ocorre nas inomogeneidades da distribuição média de matéria no Universo. Seu efeito básico é uma distorção das imagens. Finalmente, quando a separação angular entre imagens múltiplas é  pequena, da ordem de microsegundos de arco, uma situação típica para planetas,  estrelas e quasares, falamos de microlentes. Como veremos, seu efeito apesar do nome pode ser intenso.  

O artigo está organizado da seguinte forma. Na seção 2 apresentamos o desenvolvimento da área de lentes gravitacionais no que chamaremos de \textit{Idade Antiga}, finalizada em 1937 com os trabalhos de Zwicky \textbf{(Figura \ref{Fig1})}. O período de estagnação subsequente, de relativo silêncio na literatura, denominaremos de \textit{Idade Média} (1938-1962). Na seção 3,  obtemos a  equação descrevendo o caso mais simples, a chamada lente pontual,  e também algumas de suas consequências (anéis de Einstein, imagens duplas, etc.). Na seção 4,  os principais resultados teóricos e observacionais da \textit{Idade Moderna} (1963-2005) são descritos, enquanto na seção 5, narramos os contributos da \textit{Idade Contemporânea} (2006-2019). Na seção 6, discutimos brevemente um desafio futuro, a conexão entre lentes e ondas gravitacionais; uma descoberta que nos projetou na era da astronomia multi-mensageira. Finalmente, na seção 7, concluímos  o artigo tecendo as últimas considerações sobre as lentes, miragens gravitacionais e as perspectivas da área. 

\section {Lentes Gravitacionais: Idade Antiga \& Média} 
\textit{``The probability that nebulae which act as gravitational lenses will be found becomes practically a certainty" (Fritz Zwicky, 1937)}\footnote{A probabilidade de encontrar  nebulosas atuando como lentes gravitacionais torna-se praticamente uma certeza.}. 

\begin{figure}
\centering
\includegraphics[width=3.3truein,height=3.0truein]{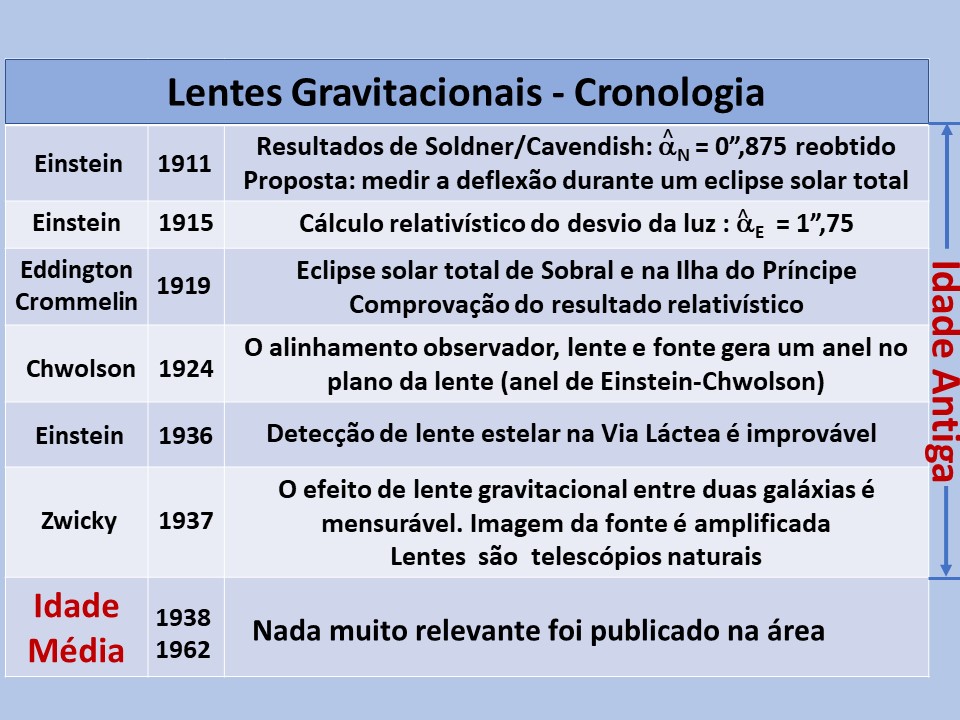}
\caption{Lentes gravitacionais (1911-1962). Linha do tempo com os principais avanços da área de LGs na Idade Antiga. Afora o experimento crucial do eclipse de 1919, todos os resultados da idade antiga são teóricos. Por completeza mostramos o período de estagnação (idade média). Somente após um silêncio de 25 anos, a área seria impulsionada por novas previsões teóricas e descobertas observacionais (secção 4).} \label{Fig1}
\end{figure}

A ideia de que a gravidade interfere na propagação da luz é uma parte razoavelmente conhecida e venerável da história da ciência.  No último quartel do século XVII, Newton tornou-se o principal defensor da teoria corpuscular da luz\footnote{Uma visão histórica da ótica newtoniana, incluindo a herança recebida dos predecessores e sua ambivalência  sobre a natureza da luz, pode ser vista em Nussenzveig \cite{MN2007} e Martins \& Silva \cite{RS2015}.}, tanto a nível experimental quanto teórico, culminando com seu famoso livro de ótica publicado em 1704 \cite{N1704}. Numa teoria corpuscular, é natural pensar que a luz, movendo-se nas proximidades de corpos celestes, seja gravitacionalmente defletida, embora Newton não tenha apresentado um cálculo explícito.

A influência da gravidade sobre o movimento da luz teve outra conexão interessante, associada aos objetos que no Século XX seriam chamados de buracos negros. John Michell (1724-1793) e Pierre Laplace (1749–1827), demonstraram que estrelas de massas suficientemente altas poderiam ser completamente escuras, ou seja, nem mesmo a luz poderia escapar da superfície \cite{M1783,L1798}. Em outras palavras, a velocidade dos corpúsculos de luz,  emitidos por um corpo esférico de massa M e raio R, poderia ser menor do que a velocidade de escape,  $v_e=\sqrt{2GM/R}$, na sua superfície\footnote{Quando a velocidade de escape é igual a velocidade da luz ($v_e=c$), temos o raio gravitacional,  $R=R_g=2GM/c^{2}$, ou raio de Schwarzschild do corpo \cite{LL1975}. Para o Sol $R_g \simeq 3$Km, mostrando que o Sol não se comporta como um buraco negro. Essa ideia seminal de buracos negros exigiria uma nova teoria de gravidade (Relatividade Geral) e quase 2 séculos seriam necessários para os "objetos invisíveis" adquirirem o presente status de corpos celestes.}.

Em 1801, Johann Soldner (1776-1833) quantificou a visão newtoniana publicando o primeiro cálculo explícito do desvio da luz nas proximidades de um corpo esférico de massa $M$ e raio $R$, supondo um corpúsculo luminoso incidindo com parâmetro de impacto $b\geq R$ \cite{S1801}.  No seu cálculo de espalhamento balístico para pequenos ângulos de deflexão, provocados por uma força $F\propto 1/r^{2}$, Soldner obteve - numa notação mais moderna - a seguinte deflexão angular\footnote{Em 1921, nos arquivos de Henry Cavendish (1731-1810), foi encontrado o mesmo resultado de Soldner, porém sem  detalhes do cálculo. Cavendish fixou a velocidade do corpúsculo $v=c$  para $r=\infty$, enquanto Soldner tomou $v=c$ para $r=R_{\odot}$. Segundo Will \cite{W1986}, as diferentes condições de contorno geram correções de segunda ordem distintas, mas os resultados coincidem em primeira ordem.}:

\begin{equation}\label{Eq1}
%\,\,\,\,\,\,\,\,\,\,\,\,\,\,\,\,\, 
{\hat{\alpha}}_N = \frac{2GM}{c^{2}b}\equiv 0'',875\, (\frac{M}{M_{\odot}}) (\frac{R_{\odot}}{b})\,.
\end{equation}
onde $G$ é a constante gravitacional de Newton, $b$ o parâmetro de impacto e $M_{\odot}=1,98\times10^{33}g$ e $R_{\odot}=6,96\times10^{10}cm$ são, respectivamente, a massa e raio do Sol.  Quando a luz defletida pelo Sol ($M=M_{\odot}$),  passando rasante a sua superfície ($b=R_{\odot}$), a expressão acima fornece o resultado newtoniano, $\hat{\alpha}_N = 0'',875$.

Em 1905, Einstein publicou  sua  teoria da relatividade especial (ou restrita) \cite{E1905}. A relatividade galileana e a dinâmica newtoniana foram generalizadas. O novo paradigma teórico  também exigia uma modifica\c{c}\~{a}o da teoria gravitacional de Newton. A no\c{c}\~{a}o de a\c{c}\~{a}o \`a dist\^{a}ncia e velocidade de propagação infinita, subjacente a visão newtoniana, era incompat\'{i}vel com as ideias de causalidade da relatividade restrita. De fato, os fundamentos da relatividade especial implicam que os  campos f\'{i}sicos devem se   propagar com velocidade finita. No mesmo ano, Henri Poincar\'e (1854-1912) também argumentou que uma descri\c{c}\~{a}o matem\'atica consistente do campo gravitacional deveria prever uma velocidade finita de propagação e a exist\^{e}ncia de radia\c{c}\~{a}o gravitacional \cite{P1905}.

Em 1911, Einstein  reiniciou sua busca pela teoria relativística da gravitação investigando a influência da gravidade sobre a propagação da luz com base no seu Princípio de Equivalência \cite{E1907}. Como resultado,  sem conhecimento prévio e de forma completamente distinta, a expressão  anteriormente deduzida por Soldner foi reobtida. No entanto,  diferente de Soldner, Einstein enfatizou a necessidade dos astrônomos tentarem medir o efeito durante um eclipse total do Sol. Em suas palavras: \textit{``Como as estrelas fixas das regiões do céu que são vizinhas do Sol se tornam visíveis quando ocorrem eclipses solares, esta consequência da teoria pode confrontar-se com a experiência"} \cite{E1911}.

\begin{figure}
  \centering
  \includegraphics[width=3.2truein,height=2.4truein]{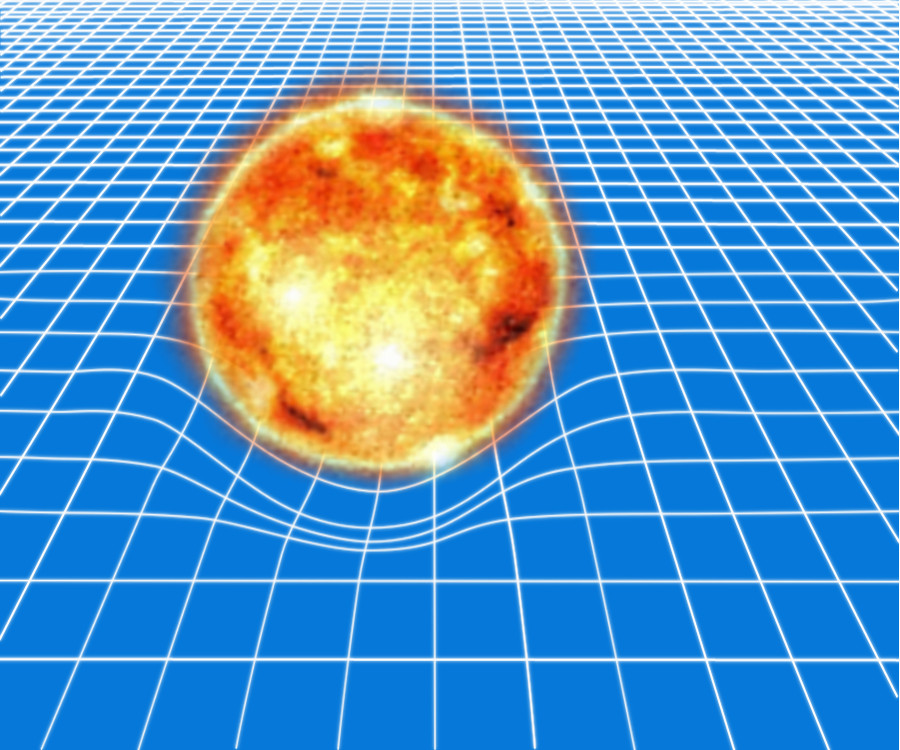}
\caption{Geometria, curvatura e matéria. Na teoria gravitacional de Einstein, a geometria do espaço-tempo depende da presença da matéria. A curvatura pode ser visualizada pelo efeito de um corpo sobre uma folha de borracha. Longe da fonte, o efeito sobre o contínuo espaço-tempo é desprezível, enquanto nas suas proximidades  torna-se curvo,  alterando o movimento de todas as partículas, inclusive da luz.} \label{Fig2}
\end{figure}

Em 1915, ao entender que a luz  propaga-se ao longo de uma geodésica, a ``linha reta" do espaço-tempo curvo,  Einstein recalculou o ângulo de deflexão, $\hat{\alpha}_E$, obtendo: 

\begin{equation}\label{Eq2}
{\hat {\alpha}}_E  = \frac{4GM}{c^{2}b}\equiv  1",75\, (\frac{M}{M_{\odot}}) (\frac{R_{\odot}}{b}).
\end{equation}
O novo resultado difere por um fator 2 do valor newtoniano original (cf. Eq. (\ref{Eq1})) e reflete a influência da curvatura do espaço-tempo (\textbf{Figura \ref{Fig2}}), provocada por uma concentração de massa nas proximidades do caminho da luz.

 Em ordem de grandeza, os dois resultados acima seguem de simples considerações dimensionais. Para um raio de luz aproximando-se de um corpo de  massa M, existem apenas duas escalas de comprimentos naturais; o raio gravitacional (Schwarszchild) $R_g= 2GM/c^{2}$ (ver \textbf{rodapé 4}) e o parâmetro de impacto $b$. O ângulo de deflexão é adimensional,  portanto, a menos de uma constante numérica ($\gamma$), podemos escrever: 
\begin{figure}
  \centering
  \includegraphics[width=3.4truein,height=1.7truein]{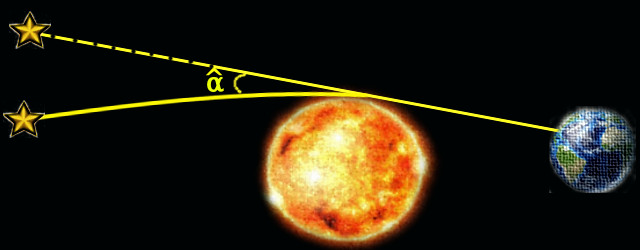}
\caption{Deflexão da luz estelar pelo Sol. A luz emitida por uma estrela sofre uma deflexão de um ângulo $\hat{\alpha}$ ao passar próximo ao Sol. O valor do ângulo depende crucialmente da teoria de gravidade adotada no seu cálculo (Newton ou Einstein) e diferem por um fator 2 a favor da teoria de Einstein, $\hat{\alpha}_E$ = 2$\hat{\alpha}_N$ (mais detalhes no texto).}\label{Fig3}
\end{figure}

\begin{equation}\label{Eq3}
\hat\alpha = \gamma\frac{R_g}{b}=\gamma\frac{2GM}{c^2 b},
\end{equation} 
onde $\gamma = 1$ no caso newtoniano e $\gamma=2$ no  relativístico.

A \textbf{Figura \ref{Fig3}} ilustra o efeito. Um raio de luz vindo de uma estrela distante sofre, ao passar próximo ao Sol,  um desvio tal que a estrela é vista deslocada da posição real. Historicamente, a deflexão da luz pelo Sol foi o primeiro exemplo observado de uma lente gravitacional. 
 
\begin{figure}\label{Medida}
\centering
\includegraphics[width=3.4truein,height=1.7truein]{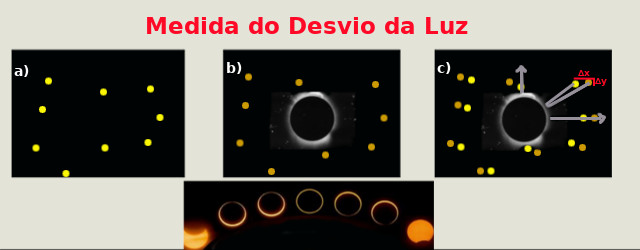}
\caption{Medida do ângulo de deflexão em 3 etapas. \textbf{a)} Estrelas são imagiadas no campo de visão sem a presença do Sol.\textbf{ b)} Posição deslocada das mesmas estrelas no momento do eclipse, e \textbf{c)} Superposição final das duas chapas fotográficas e determinação do ângulo de desvio. O resultado revelou-se de acordo com a teoria de Einstein.} \label{Fig4}
\end{figure}

 Na \textbf{Figura \ref{Fig4}}, vemos como a diferença angular entre as duas posições (aparente e real) da estrela  pode ser medida. O final do processo consiste em superpor as placas fotográficas (\textbf{Figura \ref{Fig4}c}) imagiadas na presença do sol, durante o eclipse (\textbf{Figura \ref{Fig4}b}) e algum tempo antes (ou depois), quando o Sol, no período noturno,  está fora do campo de observação (\textbf{Figura \ref{Fig4}a}). Numericamente, o efeito é muito pequeno, contudo, a diferença representada pelo fator 2 é crucial e quantifica, neste tipo de experiência, o afastamento entre as ideias de Newton e Einstein. 

É bem conhecido que a determinação
deste valor durante o eclipse solar de 1919, em Sobral e na Ilha do Príncipe (costa atlântica da África) foi a segunda confirmação observacional da Relatividade Geral. A primeira tinha sido a explicação teórica do desvio anômalo do periélio de Mercúrio, um problema que estava em aberto desde 1859 \cite{Dys20}. No entanto, podemos dizer que era um resultado para iniciados.  Por outro lado,  o caráter observacional e pedagógico envolvendo a deflexão da luz,  representou  a base da enorme popularidade adquirida por Einstein. O experimento do encurvamento da luz pelo Sol foi repetido inúmeras vezes e todas confirmaram o resultado de Einstein\footnote{Décadas depois, o resultado foi novamente confirmado com precisão melhor que $0,02$\%  utilizando técnicas interferométricas na faixa de rádio \cite{Leb95}.}.

Considerando que a relação de Einstein (\ref{Eq2}) foi confirmada em 1919, existem duas  consequências teóricas imediatas para o fenômeno de lentes gravitacionais: (i) a equação pode ser utilizada de forma inversa, ou seja, se o raio do objeto defletor é conhecido, a deflexão medida implica na determinação de sua massa, e (ii) a determinação da massa independe da natureza da matéria constituinte [seja matéria bariônica (nêutrons, prótons, elétrons, etc.)], ou mesmo  um tipo de matéria ainda desconhecida. A partir da década de 80, tais efeitos se revelariam importantes no domínio cosmológico.

Em 1924, Chwolson \cite{C1924} argumentou  que no caso de alinhamento perfeito entre observador, lente e fonte, a imagem da estrela mais distante (fonte) seria vista como um anel em torno da lente, a estrela mais próxima. Esse efeito foi posteriormente denominado anel de Einstein\footnote{Um leitor mais atento perguntaria:  \textit{por que não anel de Chwolson?} Em 1997, encontrou-se  um caderno de anotações de Einstein com o cálculo do anel feito em 1912. Como seria esperado, o resultado continha o erro do fator 2, oriundo da deflexão obtida em 1911. O artigo de Einstein \cite{E1936} é baseado nas antigas notas \cite{Renn1997}. O nome anel de Einstein-Chwolson (ainda) não tornou-se popular.}. 

Em 1936, Einstein \cite{E1936} fez o cálculo do anel e concluiu também que seria improvável a observação do encurvamente por estrelas mais próximas, da luz  emitida por estrelas  distantes dentro da nossa galáxia.  

Em 1937, Fritz Zwicky (1898-1974)  publicou  dois artigos memoráveis, ambos de meia página  \cite{Z1937a,Z1937b}. No primeiro, ele argumentou que as nebulosas (o nome das galáxias na época), seriam mais adequadas do que estrelas para observações do fenômeno de lentes gravitacionais, o que seria até natural, pois o ângulo de deflexão é proporcional a massa. Ressaltou também o interesse da pesquisa por outras razões, dentre elas: (i) as galáxias oferecem a possibilidade de um teste adicional para a relatividade geral, (ii) o efeito de  lente gravitacional seria amplificado mais de uma ordem de magnitude, quando uma galáxia inteira desempenhasse o papel de lente para outra ainda mais distante, tornando sua detecção bastante provável, e  (iii) o efeito forneceria uma determinação  direta da massa da galáxia lente e poderia sugerir novas linhas de investigação para os problemas cosmológicos. 

Os resultados de Zwicky não tiveram influência imediata, ou seja, não produziram eco na literatura.  Transcorreu um longo e silencioso período sem descobertas relevantes na área\footnote{Provavelmente, isto aconteceu devido  a 2a. guerra mundial e a concentração de grande parte da comunidade em áreas também emergentes e de interesse teórico, econômico e militar, notadamente, física  nuclear, plasmas, partículas e suas aplicações.}, definindo neste artigo a \textbf{Idade Média} das Lentes Gravitacionais (ver \textbf {Figura \ref{Fig1}}). 

\begin{figure}
\centering
\includegraphics[width=3.3truein,height=2.6truein]{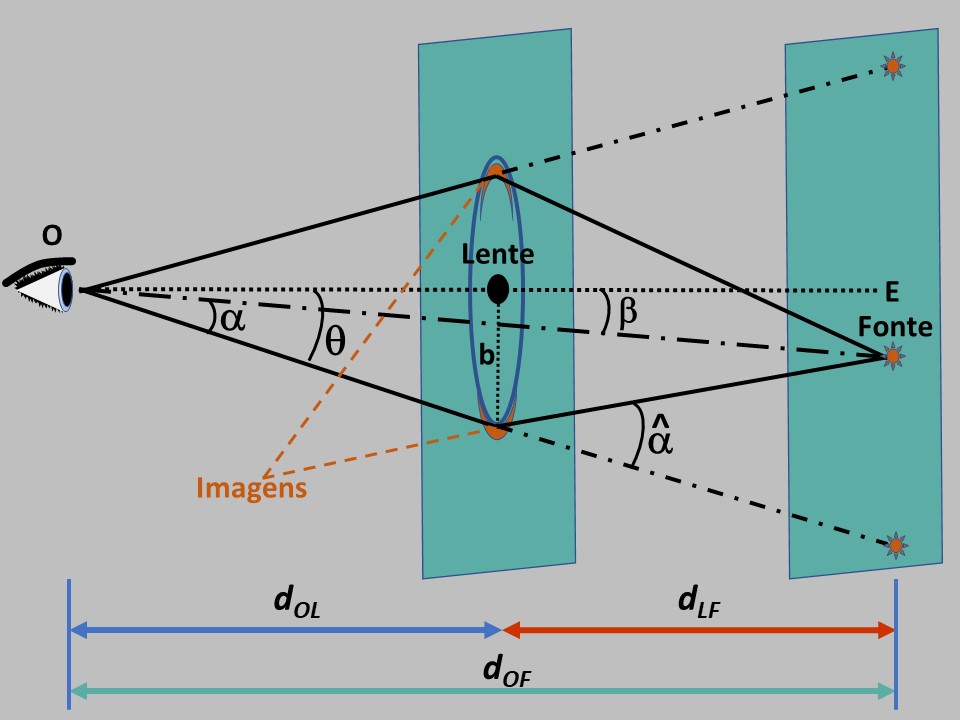}
\caption{\textit{Geometria para  eventos de lente pontual.} As respectivas distâncias entre observador (O), lente (L) e fonte (F) são $d_{OL}$, $d_{LF}$ e $d_{OF}$.  A luz emitida pela fonte é defletida próxima a lente. Os ângulos $\alpha$,  $\theta$ e $\beta$  são pequenos e  medidos em relação ao eixo ótico (\textbf{EO}). Note que $\alpha$ é o ângulo entre a fonte e uma das imagens, enquanto $\beta$ e $\theta$ especificam, respectivamente, a posição da fonte no céu e da imagem inferior. O ângulo $\hat{\alpha}$ é a deflexão angular (valor de Einstein) e $\text{b}$ o parâmetro de impacto. Para $\beta\neq 0$, duas imagens (distorcidas) são formadas, mas o caso $\beta=0$ (alinhamento total) é degenerado, com a imagem formando o chamado anel de Einstein.} 
\label{Fig5}
\end{figure}

 Os artigos seminais de Zwicky  seriam retomados somente a partir dos anos 60, com a descoberta dos quasares e novas propostas teóricas. No entanto, antes de continuar nossa cronologia, apresentaremos alguns conceitos básicos da teoria de lentes gravitacionais. Como veremos a seguir, os resultados são todos de natureza puramente geométrica. Além de simples, serão úteis para entender, nas seções seguintes, os principais contributos da Idade Moderna e Contemporânea.

\section{Equação da Lente, Anel de Einstein, Imagens  Duplas, Múltiplas  e Magnificação}

 O exemplo mais antigo e pedagógico de lenteamento gravitacional ocorre quando a lente (e a fonte) podem ser tratadas como objetos pontuais. A fonte pontual (ou lente de Schwarzschild) é um caso particular de simetria circular no plano da lente, como ocorre, por exemplo,  numa galáxia elíptica. A descrição geral é válida para campos gravitacionais estacionários fracos, pequenos desvios angulares e quando o tamanho típico do defletor é muito menor do que o meio de propagação (lente fina). A lente  pontual discutida a seguir é uma manifestação do que é usualmente chamado de microlente.

 \subsection{Equação da Lente}

 Considere uma fonte de luz (F) bem atrás de uma lente pontual  (L) de massa M. Para um observador (O), a luz defletida poderá seguir dois caminhos distintos ou no caso "geometricamente  degenerado"\, formar um anel. 

Na \textbf{Figura \ref{Fig5}} ilustramos  a geometria canônica
 de uma microlente pontual (estática) na chamada aproximação da ótica geométrica\footnote{Na aproximação da ótica geométrica, o comprimento de onda da radiação deve ser bem menor do que todas as escalas macroscópicas envolvidas no processo ($\lambda <\,< L$).}. Em geral, os ângulos são tão pequenos que as vezes não é possível resolver a fonte distante por trás da lente, embora as imagens (miragens) distorcidas sejam mais facilmente visíveis (as imagens tem separações angulares menores do que 30 segundos de arco).  Os resultados a seguir permanecem válidos mesmo quando o sistema está tão distante que o modelo cosmológico padrão ($\Lambda$CDM) precisa ser considerado\footnote{A única diferença é que devido a expansão universal, as distâncias envolvidas, chamadas de distâncias de diâmetro angular ($d_{A}$), devem levar em conta a presente taxa de expansão do Universo (constante de Hubble, $H_0$), o conteúdo material ($\Omega_{M}+\Omega_{\Lambda}=1$) e o desvio para o vermelho (\textit{redshift z}) das fontes distantes\cite{LA2000,Santos2007,SL2008,LS2018}.}.
 
O ângulo  $\hat{\alpha}$ é definido pelo valor de Einstein (ver Eq. (2)) e claramente depende do parâmetro de impacto $b$ ou, equivalentemente, do ângulo $\theta$ 
\begin{equation}
\label{Eq4}
\hat{\alpha} =\frac{4GM}{c^2 b}.
\end{equation}
Considerando que o ângulo $\beta= \theta - \alpha$ (ver \textbf{Figura \ref{Fig5}}) obtemos a chamada Equação da Lente: 

\begin{equation}\label{Eq5}
\beta  = \theta -\hat{\alpha}\frac{d_{LF}}{d_{OF}}.
\end{equation}
onde utilizamos o resultado válido  para pequenos ângulos, $\alpha = \hat{\alpha} d_{LF}/d_{OF}$.  Note também  que o  parâmetro de impacto é dado por $b= d_{OL}\theta$. Substituindo em (3) e inserindo o resultado na expressão (4), a equação da lente pontual pode ser reescrita como:

\begin{equation}
\label{Eq6}
\beta = \theta - \frac{\theta^2_E}{\theta},   
\end{equation}
onde definimos, 
\begin{equation}\label{Eq7}
\theta^2_E=\frac{d_{LF}}{d_{OF} d_{OL}}\frac{4GM}{c^2}.
\end{equation}
 Como veremos, o valor de $\theta_E$ está associado a uma escala característica fundamental na teoria de lentes. 
 
\subsection{Anel de Einstein}

Quando observador, lente e fonte, estão alinhados no eixo ótico \textbf{OE} temos uma situação curiosa. Neste caso $\beta = 0$ (ver \textbf{Figura \ref{Fig5}}). As duas possíveis soluções  de \eqref{Eq6}  são degeneradas, no seguinte sentido: devido a simetria rotacional, a imagem é vista como um anel, bastante referido como anel de Einstein. Sua abertura angular medida do centro da lente é dado por:  
\begin{equation}\label{Eq8}
\theta_E = \sqrt{{\frac{d_{LF}}{d_{OF}d_{OL}}\frac{4GM}{c^2}}}=    \sqrt{\frac{4GM}{c^2D}}. %0".9\left(\frac{M}{M_{odot}}\right)^{1/2}\left(\frac{R_{odot}{D}\right)^{1/2}\,\,.\,\,\,\,\,\,\,
\end{equation}

\begin{figure}
\centering
\includegraphics[width=3.3truein,height=2.6truein]{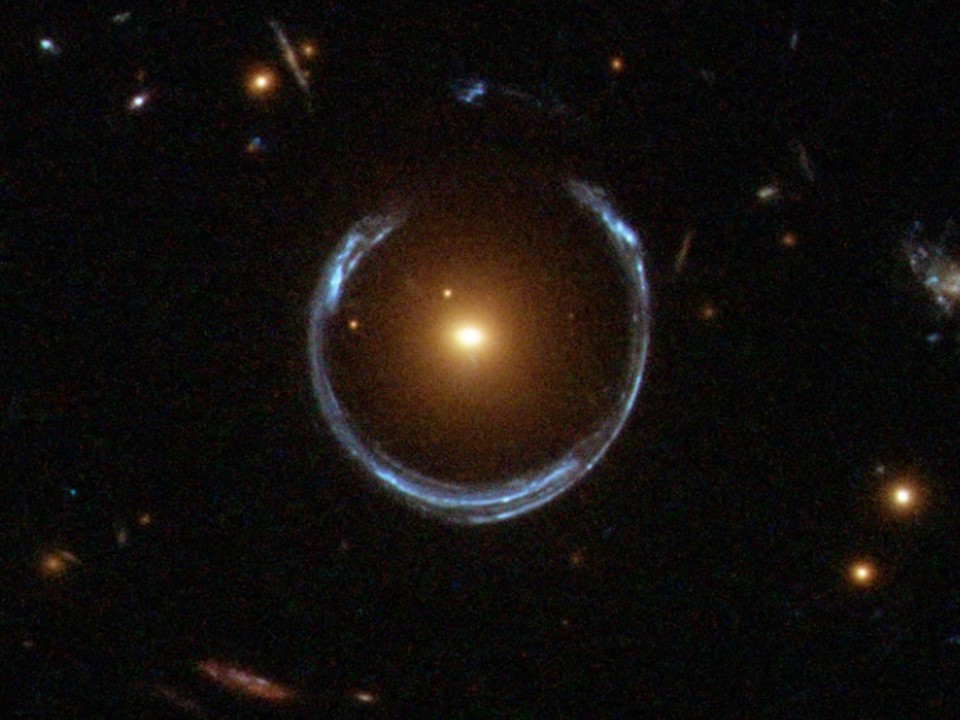}
\caption{Formação de um anel de Einstein galáctico, como sugerido por Zwicky. A imagem de uma galáxia azul mais distante (fonte) foi distorcida por uma galáxia elíptica mais próxima (lente). Devido a precisão do alinhamento a galáxia de fundo é distorcida  formando  um anel quase completo (ver texto). Suas propriedades permitem  determinar o conteúdo de massa da galáxia, incluindo a matéria escura (crédito da imagem: ESA/Hubble \& NASA).} 
\label{Fig6}
\end{figure}

 O valor de  $\theta_E$  depende  de uma combinação de todas as distâncias envolvidas e não apenas da distância lente-fonte. A quantidade, $D = (d_{OL}{d_{OF}})/d_{LF}$, fornece a escala de distância efetiva do  processo. Note também que para uma lente esférica alinhada basta considerar a distribuição de matéria (perfil de densidade),  cuja massa total é $M=M(\theta_E)$.
%As equações (\ref{EL}-\ref{EL1}) podem também descrever uma fonte extensa se a simetria esférica é mantida, como no caso de uma galáxia elíptica. Neste caso,  basta  considerar que M=M($\theta$).

Como foi visto, o anel de Einstein, $\theta_E \propto (M/D)^{1/2}$. Logo, introduzindo $M_{\odot}$ e uma distância típica (Via Láctea), D=10Kpc, temos: $\theta_E \simeq (10^{-3})''(\frac{M}{M_{\odot}})^{1/2}(\frac{D}{10kpc})^{-1/2}$. Para uma galáxia elíptica de $M= 10^{12}M_{\odot}$, a uma distância efetiva $D=1Gpc$, obtemos $\theta_E \simeq 1"$, o que mostra ser bem mais fácil o lenteamento de galáxias.

Na \textbf{Figura \ref{Fig6}}, mostramos um exemplo do anel de Einstein formado por uma galáxia elíptica.  Casos mais complexos envolvendo  distribuições finitas de matéria  sem simetria, são também  analiticamente tratáveis com as devidas integrações (na lente, fonte ou em ambos)\footnote{Num certo sentido, a lente pontual está para o caso geral de lentes, assim como  o campo elétrico da carga pontual está para os campos de distribuições finitas de cargas. Ou seja, casos mais gerais podem ser obtidos por integrações (na fonte e na lente) e/ou no aumento da dimensionalidade do problema \cite{SEF1999,Mol2002}.}.

\subsubsection{Imagens Duplas}
A equação da lente  pontual (ou circular) descrita por \eqref{Eq6}, pode ser reescrita como
\begin{equation}\label{Eq9}
\theta^2- \beta\theta - \theta^2_E=0\,,
\end{equation}
cujas soluções correspondem a duas imagens:
 
 \begin{equation}
\label{Eq10}
\theta_{\pm}=\frac{1}{2}\left(\beta \pm\sqrt{\beta^2+4\theta_E^2}\right).
\end{equation}
Note que as duas imagens estão em lados opostos, mas não são simétricas.  Das soluções acima e também da \textbf{Figura \ref{Fig5}} vemos que uma das imagens (superior)  está dentro do anel de Einstein e a inferior fora. Tal assimetria é gerada pelo  posicionamento da fonte (se está acima ou abaixo do eixo ótico). Se a fonte estivesse acima, seria a imagem inferior que estaria dentro do raio de Einstein ($\theta < \theta_E$) e vice-versa. Portanto, uma das imagens está sempre dentro e a outra fora do anel de Einstein. Observe também que a separação entre as imagens 

\begin{equation}\label{Eq11}
 \Delta\theta_{\pm} = \theta_{+} - \theta_{-} = \sqrt{4\theta_E^2 + \beta^2} \geq 2\theta_E\, ,   
\end{equation} 
tem seu valor mínimo determinado por $\theta_E$. Se $\beta$ cresce a separação aumenta.

\subsection{Magnificação} 

O lenteamento gravitacional preserva o brilho superficial da fonte, pois o número de fótons emitidos é conservado. Em outras palavras, uma LG apenas redistribui os fótons emitidos pela fonte. No entanto, o ângulo sólido da fonte pode ser alterado pelo efeito de lente e, portanto, o mesmo acontece com o fluxo total. A variação do fluxo total depende da razão entre os ângulos sólidos de cada imagem e da fonte. Como resultado, podemos definir a magnificação (${\mathcal{M}}$), e também o seu oposto, a demagnificação ou enfraquecimento, como uma razão entre áreas \cite{NB1996}:

\begin{equation}
%Magnifica\c{c}\~a{o} MAG \equiv 
\mathcal{M} = \frac{Area \,\, da \,\,Imagem}{Area\,\, da\,\, Fonte}.
\end{equation}
É possível também mostrar que a razão acima para lentes pontuais implica  que a magnificação de cada imagem pode ser escrita como (ver Apêndice B)

\begin{equation}\label{13}
{\mathcal M_{\pm}}=\frac{\theta_{\pm}}{2\beta}
\left(\frac{\beta}{\sqrt{\beta^{2} + 4\theta^{2}_E}\pm 1} \right),
\end{equation}
onde $\theta_{\pm}$ são as soluções das imagens na equação \ref{Eq10}. Finalmente, introduzindo a quantidade normalizada,  $x=\beta/\theta_E$, segue que:

\begin{equation} \label{Eq14}
{\mathcal M_{\pm}}=\frac{x^{2} + 2}{2x\sqrt{x^{2} +4}} \pm \frac{1}{2},
\end{equation}
e somando as duas expressões acima, temos a magnificação total:

\begin{equation}\label{Eq15}
{\mathcal M} = {\mathcal M_{+}} + {\mathcal M_{-}} = \frac{x^{2} + 2}{x\sqrt{x^{2}+4}}.
\end{equation}
Como uma ilustração, considere uma fonte sobre o raio de Einstein, ou seja,  $\beta=\theta_E$. Neste caso,  temos $x=1$ e uma magnificação total de ${\mathcal M}=1,34$. Vários exemplos  de magnificação (demagnificação) foram detectados até o presente.  

\subsection{Lentes como Telescópios Naturais}

Astrônomos costumam também dizer que telescópios são máquinas do tempo, pois permitem saber como eram os objetos ou mesmo o Universo no passado. No mesmo sentido, as lentes gravitacionais funcionam como verdadeiros telescópios naturais. Isto ocorre  devido ao efeito de magnificação, pois as imagens  parecem mais brilhantes do que a fonte. Em muitos casos, a  magnificação  possibilita a detecção de uma fonte distante.  Proto-galáxias em altos redshifts tem sido descobertas e seu espectro revelado por efeitos de lente forte agindo como um "telescópio" natural. Um efeito que pode também  explicar porque quasares com altos \textit{redshift}s podem ser encontrados próximos a galáxias de baixos \textit{redshifts}.

\begin{figure}
\centering
\includegraphics[width=3.2truein,height=3.2truein]{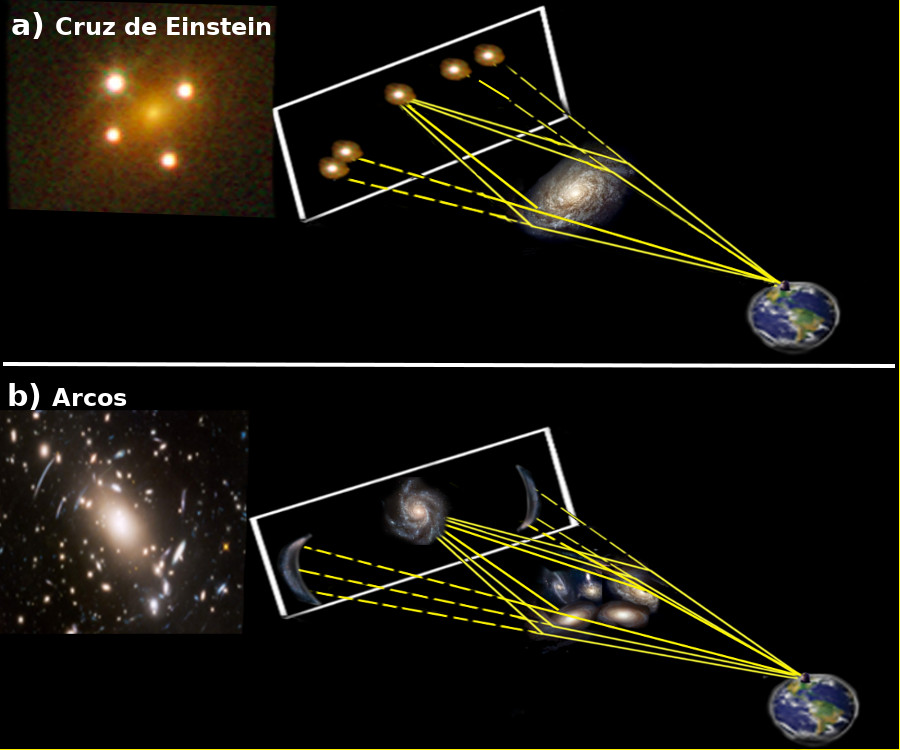}
\caption{Imagens múltiplas. Quando existem vários caminhos óticos independentes, diferentes imagens são observadas. \textbf{a)} Cruz de Einstein; \textbf{b)} Formação de arcos (gigantes e pequenos) em aglomerados.} \label{Fig7}
\end{figure}

Em 2017, foi estabelecido um novo recorde de magnificação: uma galáxia distante, foi imagiada pelo Hubble Space Telescope (HST) graças a um aumento de 30 vezes pelo efeito de lente gravitacional \cite{LNT2017}. 

\subsection{Imagens Múltiplas e Critérios na Identificação do Lenteamento}

 Como foi visto, numa lente pontual ou circularmente simétrica, formam-se duas imagens ou um anel de Einstein.  No caso geral, assimétrico, ou na presença de um conjunto de lentes,  o tratamento matemático é mais complexo, pois a equação escalar da lente (\ref{Eq5}) torna-se vetorial \cite{SEF1999c,Mol2002b}. O tratamento matemático do caso geral está fora do escopo do presente artigo, embora seja  simples de entender fisicamente a formação de imagens múltiplas.

Na \textbf{Figura \ref{Fig7}} mostramos uma visão qualitativa da formação de várias imagens. Quatro ou mais miragens podem ser formadas numa lente assimétrica ou constituída por muitas lentes quase pontuais. A figura é ilustrada com a chamada cruz de Einstein, um conjunto de 4 miragens\footnote{Um teorema devido a William Burke (1941-1996), baseado no estudo da formação de caústicas e linhas críticas do lenteamento,  demonstrou que o número de imagens múltiplas é sempre ímpar \cite{B1981}; embora uma das imagens sendo demagnificada não seja vista, tal como ocorre na cruz de Einstein (ver também \cite{M1985}).} observada \textbf{Figura \ref{Fig7}a}. No caso de muitas fontes independentes, como por exemplo, num aglomerado de galáxias, temos também formação de arcos gigantes ou de pequenos arcos (\textbf{Figura \ref{Fig7}b}). As primeiras imagens de arcos gigantes (lentes fortes) foram obtidas  por Lynds e Petrosian \cite{LP1986}, enquanto os pequenos arcos foram vistos primeiramente  por Tyson et al. 1990 \cite{T1990}. Na seção 4.2 discutiremos brevemente lentes fortes e fracas.

Por outro lado, para os objetos observados serem classificados como imagens, alguns critérios básicos devem ser obedecidos \cite{SEF1999b}. Os mais importantes são: (i) As 2 imagens devem estar bem próximas no céu, (ii) A razão e a forma dos fluxos nas diferentes bandas espectrais devem ser as mesmas,  (iii) No caso cosmológico, as imagens devem ter o mesmo desvio para o vermelho (\textit{redshift}), (iv) Uma possível lente deve estar na vizinhança das imagens, e (v) as variações temporais nas diferentes imagens (devido a variações na fonte) devem estar correlacionadas (\textit{time-delay}). Retornemos agora a nossa cronologia das lentes gravitacionais.

\section {Idade Moderna}

\begin{figure}
\centering
\includegraphics[width=3.3truein,height=3.0truein]{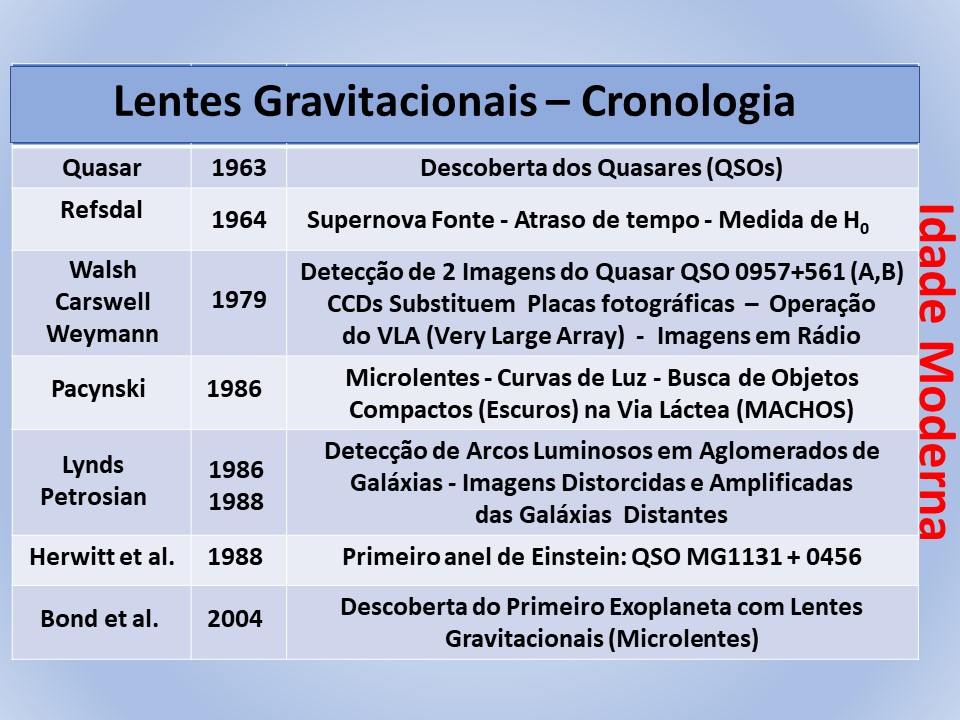}
\caption{Lentes gravitacionais na Idade Moderna (1963-2005). Linha do tempo mostrando os principais avanços teóricos e observacionais na área de lentes gravitacionais no período. A identificação dos quasares (1963) e a  primeira observação de uma imagem dupla em 1979 (QSO 0957+561 A \& B) estão entre as mais extraordinárias descobertas do período.}\label{Fig8}
\end{figure}

Em 1963 foram  descobertos os quasares\footnote{https://pt.wikipedia.org/wiki/Quasar, consultado em 15/07/2019.}, uma classe de objetos com aparência quase estelar (quasistellar object, QSO no acrônimo inglês). Os quasares \cite{Schmidt1963} são extremamente compactos, luminosos e  estão localizados a grandes distâncias, ou seja, em \text{redshifts} mais altos do que a maioria das galáxias. Imediatamente percebeu-se que os QSOs formam uma população especial de fontes pontuais distantes, convenientes para formação de imagens (ou miragens) como sugerido por Zwicky. Sua luz certamente seria lenteada  pelas galáxias e aglomerados na longa jornada até a Terra. A identificação dos quasares\footnote{QSOs estão entre os eventos mais energéticos do Universo. São capazes de emitir centenas ou milhares de vezes a energia eletromagnética de nossa galáxia. Sua aparência e potência são explicadas por um superburaco negro acretando matéria da vizinhança. Atualmente, já foram observados QSOs com \textit {redshifts} $z>6$.} revelou-se fundamental no estudo das lentes gravitacionais e demarcarão nesse texto o prelúdio da era moderna (ver \textbf{Figura \ref{Fig8}}).

Em 1964, Refsdal \cite{Re1964a,Re1964b} analisou o lenteamento por uma galáxia supondo que a fonte pontual de luz era uma supernova (SN) distante.
Sendo a SN um evento transiente, a luz desviada por uma lente galáctica (situada entre a SN e a Terra), permitiria  que a explosão fosse vista em momentos distintos nas diferentes imagens\footnote{A partir de 1998, a experiência adquirida na busca de SNs em altos \textit{redshifts} no contexto cosmológico, criou as condições  adequadas para o uso de SN como fonte no processo de lente. O lenteamento de uma SN foi finalmente observado em 2014, com a descoberta da chamada Supernova de Refsdal contendo várias imagens (ver \textbf{Figura \ref{Fig18}}).}. Refsdal calculou o atraso de tempo ("time delay"), devido a diferença de percurso entre a fonte e as imagens. Isso ocorreria não apenas com SNs, mas também com qualquer fonte pulsante ou variável.  Foi também argumentado que além de uma estimativa da massa da galáxia defletora, as miragens formadas determinariam a constante de Hubble - $H_0$ - combinando os parâmetros observados no sistema lente-fonte e no atraso de tempo. Na idade contemporânea, veremos como  essa notável previsão de Refsdal poderá resolver a tensão existente nas medidas da constante de Hubble ($H_0$) no contexto cosmológico (seção 5.5). 

\begin{figure}
\centering
\includegraphics[width=3.3truein,height=2.5truein]{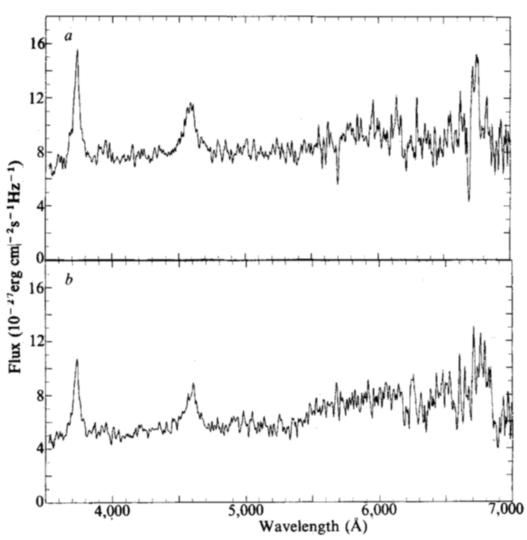}
\caption{Espectros dos QSOs $0957+561, A\, e\, B\,\, \cite{WCW1979}$. Note a semelhança dos espectros extraídos no mesmo redshift.  Espectros mais precisos em várias bandas, revelaram que os critérios gerais para identificação de miragens, formulados no final da seção 3.3, são obedecidos nos quasares gêmeos.} \label{Fig9}
\end{figure}

Em 1979,  Walsh, Carswell \& Weymann \cite{WCW1979} observaram a primeira formação de duas imagens por uma fonte extragaláctica, o quasar $QSO\,\,0957+561 (A\,\, \&  \,\,B)$ \cite{WCW1979}. Embora previamente selecionado de um catálogo de rádio fontes, as duas imagens do quasar também aparecem no ótico. A similaridade dos espectros extraído das imagens no \textit{redshift} $z\simeq 1,4$,  mostram que são miragens de um único objeto lenteado (\textbf{Figura \ref{Fig9}}).

\begin{figure}
\centering
\includegraphics[width=3.2truein,height=2.5truein]{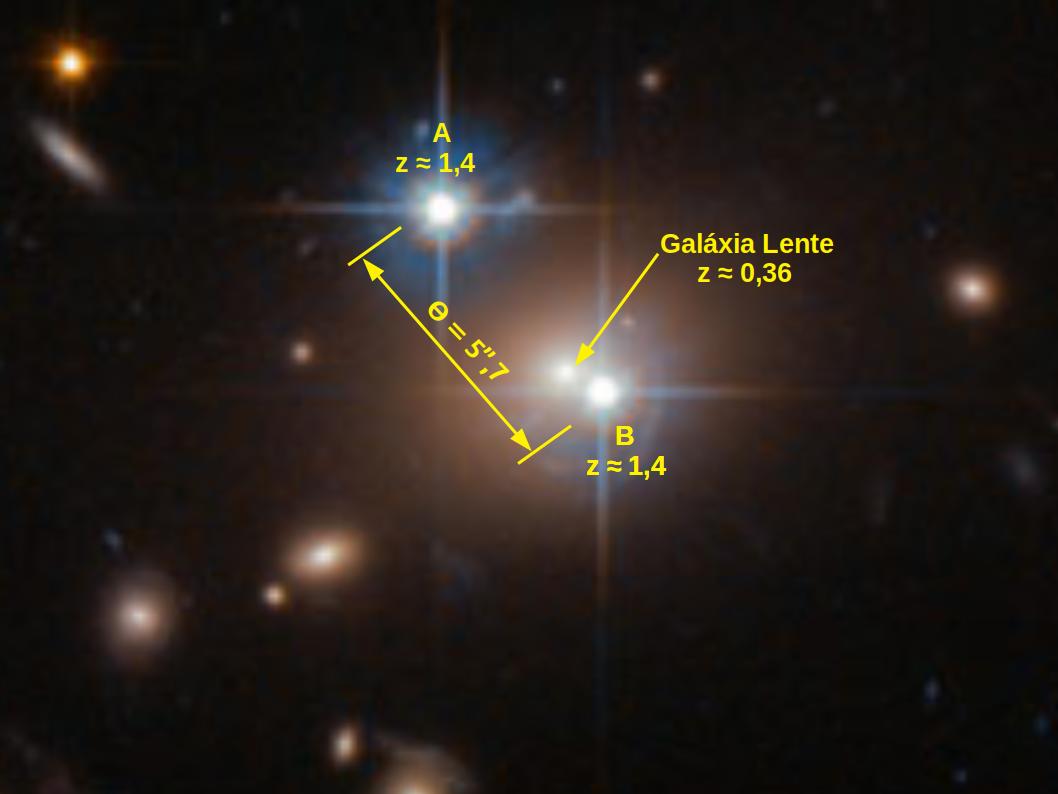}
\caption{Imagem do telescópio espacial Hubble (HST), mostrando a dupla miragem do quasar QSO $0957 + 561$ (os 2 pontos mais brilhantes, A e B). A galáxia lente (YGKOW G1), está mais próxima de B (mais detalhes no texto). Crédito da Imagem: ESA/Hubble \& NASA (com adaptação pedagógica).} \label{Fig10}
\end{figure}

Na \textbf{Figura \ref{Fig10}}, mostramos uma imagem mais recente dos gêmeos QSO 0957+561 (A\,\,\& \,\,B), incluindo sua galáxia lente (YGKOW G1), situada no \textit{redshift} $z \approx 0,36$. As imagens são os pontos mais brilhantes, separados por 5,7 segundos de arco. A lente situa-se entre os dois,  estando  mais próxima de B. Diversos estudos posteriores mostraram que os critérios gerais para identificação de imagens, enumerados no final da seção 3.3, são obedecidos, a menos do último pois a fonte não é variável. Provavelmente, o quasar QSO 0957+561 (e suas miragens), é a fonte mais estudada da astronomia moderna.

Naturalmente, a descoberta do quasar duplo atraiu atenção geral. Além de estimular as pesquisas em LGs,  mostrou o interesse do imagiamento de objetos brilhantes e longínquos do Universo, tais como os quasares, supernovas e as primeiras gal\'{a}xias. O sucesso foi também possível devido a uma profunda mudança tecnológica: a substituição das antigas placas fotográficas por CCDs, produzindo um aumento considerável de sensibilidade na detecção de fótons\footnote{Em 1979, a entrada em operação do rádio interferômetro VLA (\textit{Very large Array}), mostrou que os dois quasares são fontes de rádio compactas de mesmo espectro. Em 1980, o uso do VLA também permitiu a identificação da galáxia lente \cite{Y1980} (ver \textbf{Figura \ref{Fig10}}).}.

Neste ponto é importante também mencionar que o primeiro anel de Einstein, a fonte de rádio MG1131 + 0456,  foi descoberto em 1988 por Hewitt e colaboradores \cite{Hewitt1988}, utilizando também o VLA (\textbf{Figura \ref{Fig8}}).
%p

\subsection{Lentes Gravitacionais e  Matéria Escura}

Supondo que a matéria dominante é bariônica ou normal\footnote{A matéria existente em planetas, estrelas, etc. Tecnicamente, são partículas (pesadas) do modelo padrão da física de partículas elementares.}, a chamada curva de rotação das galáxias revelaram um paradoxo inesperado. 

Para entender o mistério,  considere um objeto girante de massa $m$ a uma distância $r$ do centro de uma galáxia aproximadamente esferoidal. A massa da Galáxia é $M(r)$ e a força gravitacional sobre o objeto girante é a força centrípeta ($F_G=F_C$). Newtoniamente, a velocidade do objeto é dada por $v(r) = \sqrt{GM(r)/r}$. 

Fazendo observações além da parte luminosa, teríamos $M(r)=constante \Rightarrow\,\,\,v\propto r^{-\frac{1}{2}}$, ou seja, a velocidade tangencial obedeceria uma lei kepleriana, tal como ocorre com os planetas no caso do Sol. 

\begin{figure}
\centering
\includegraphics[width=3.3truein,height=2.7truein]{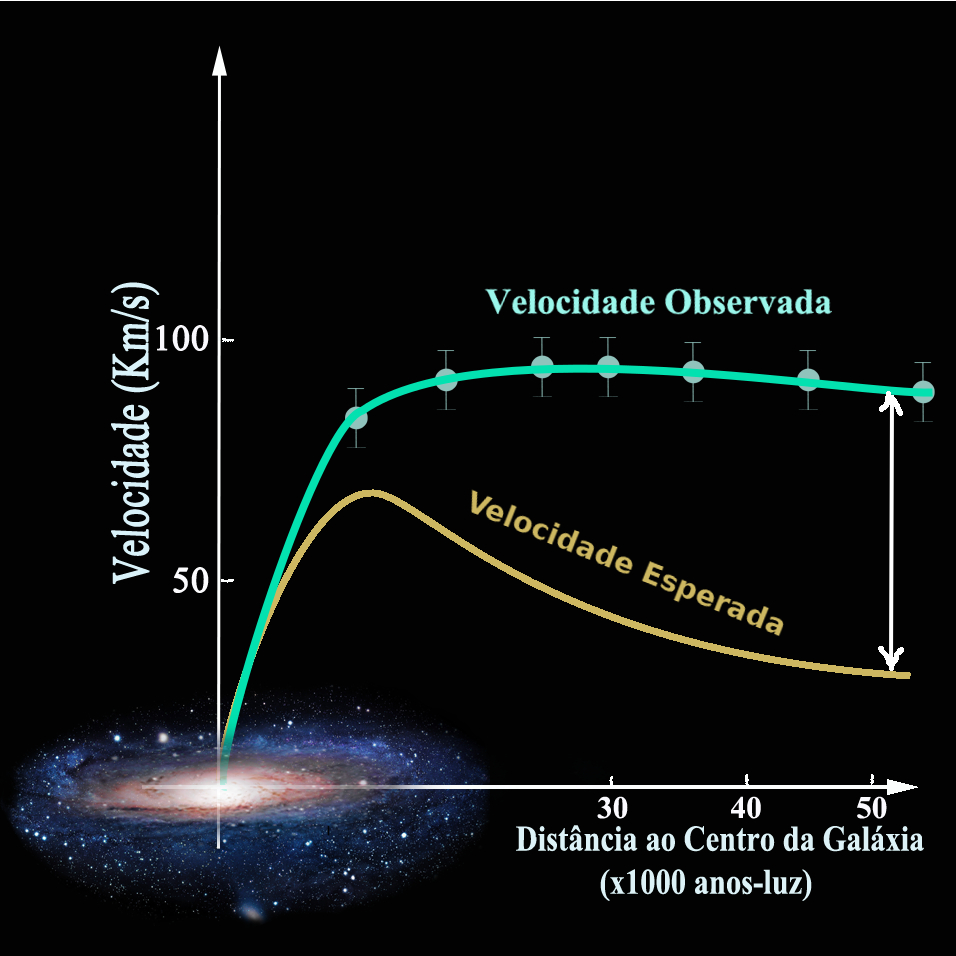}
\caption{Curva de rotação de uma galáxia espiral. A diferença entre a curva superior (onde estão os dados) e a inferior teórica pode ser explicada pela presença de uma componente invisível que interage apenas gravitacionalmente (ver texto)
.} \label{Fig11}
\end{figure}

No entanto, as observações mostraram que as velocidades na periferia não diminuíam, ficando aproximadamente constante,  diferente do previsto  newtoniamente. Uma solução possível é a existência de uma forma de matéria invísivel,  ou seja, que interage gravitacionalmente, mas não brilha. Se $M(r) \propto r$,\, teremos $v \simeq const.$, resolvendo o problema (ver \textbf{Figura \ref{Fig11}}).

No início dos anos 70, Kenneth Freeman \cite{KF70}, Vera Rubin e colaboradores \cite{VR,VR2}, sugeriram que a existência da matéria escura fria (\textit{cold dark matter}, CDM no acrônimo inglês) seria a explicação mais simples para as curvas de rotação  das galáxias espirais. Ostriker e Peebles \cite{OP73} já haviam argumentado que um halo esférico de matéria escura nas galáxias espirais poderia também garantir a estabilidade do disco luminoso. 

\begin{figure}
\centering
\includegraphics[width=3.3truein,height=2.7truein]{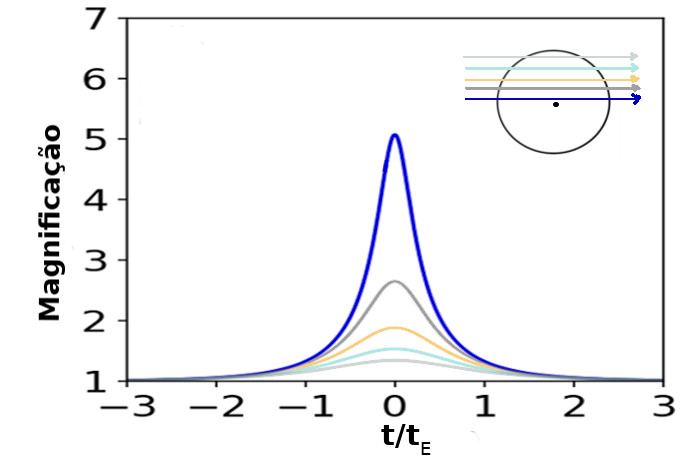}
\caption{Variação da magnificação (curva de luz) como função do tempo em lentes pontuais.  As curvas foram traçadas para 5 parâmetros de impacto diferentes (não mostrados). O circulo é o anel de Einstein e as setas indicam as trajetórias. Quanto menor o parâmetro de impacto, maior a magnificação.  Note a simetria da curva em torno de t=0. (Adaptado de  \cite{BP1996}).} \label{Fig12}
\end{figure}

A conexão matéria escura, lentes gravitacionais e cosmologia foi imediata. Dois tipos de candidatos foram propostos:  (i) objetos compactos (buracos negros, estrelas de nêutrons, anãs marrons, planetas)  e (ii) um zoológico de partículas elementares  estáveis, fósseis do Universo primitivo, tais como, o Higgs, Axions, Neutralinos e WIMPs\,\footnote{Weekling Interacting Massive Particles.}. As observações revelariam sua abundância cósmica, natureza,  ou descartaria candidatos. Estava  aberta a era da chamada "astronomia da matéria invisível", uma inesperada  oportunidade para as LGs. Veremos a seguir como  as LGs descartaram a  proposta de objetos compactos como formadores dos halos escuros\footnote{A natureza do possível  candidato a matéria escura, oriundo da física de partículas,  ainda permanece indeterminada \cite{P1993,DM2019}.}.     
Em 1986, Bohdan Paczynski (1940-2007), sugeriu que a variação da magnificação devido a efeitos cinemáticos em microlentes\footnote{Paczynski foi o primeiro a utilizar o termo microlente \cite{W2006b}.}, poderiam identificar objetos compactos escuros no halo da Via Láctea (lentes), pois seriam iluminados por estrelas das nuvens de Magalhães \cite{BP1986a,BP1986b}. Os objetos compactos cruzando a linha de visada das estrelas provocaria uma variação temporária do  brilho (aumento seguido de um decaimento na magnificação)  que poderia ser monitorado. 

\begin{figure}
\centering
\includegraphics[width=3.3truein,height=2.7truein]{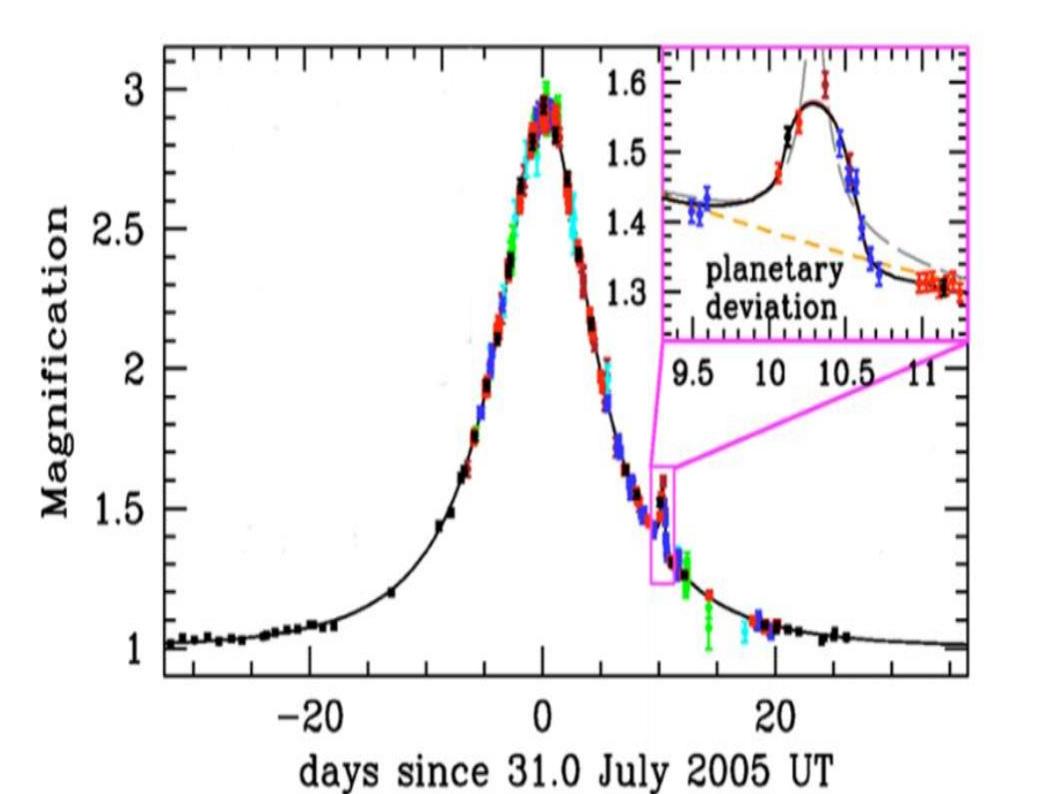}
\caption{Curva de luz de um efeito de microlente planetária $OGLE-2005-BLG-390$. As diferentes cores representam as contribuições de vários telescópios. O canto superior é um \textit{"zoom"} da  anomalia  planetária, cujo  melhor ajuste é indicado pela linha preta sólida. O planeta tem uma massa de $\simeq 5,5 M_{terra}$ (ver texto) e está a uma distância de 2,6 UA da estrela hospedeira (extraída de \cite{Beaulieu2006}).} \label{Fig13}
\end{figure}

Várias colaborações/projetos foram organizados com esses objetivos, tais como, MACHOs\footnote{MACHOs\,\, $\equiv$ Massive\,\, Compact\,\, Halo\,\,\,Objects}, EROS\footnote{EROS\,$\equiv$ Expérience\,\, pour\,\, la\,\,Recherche\,\,d'Objets\,\, Sombres} e  OGLE\footnote{OGLE\,\,$\equiv$\, Optical \,\,Gravitational\,\, Lensing\,\,Experiment}. A variação da magnificação com o tempo é medida pela ``curva de luz" de Paczynski (\textbf{Figuras \ref{Fig12} e \ref{Fig13}}), e sua observação permite estimar a massa da lente. \textit{Como funciona o método?} 

A duração do evento é fixada pelo tempo característico de Einstein. Supondo uma  velocidade transversa da lente constante temos, $t_E=r_E/V_{\perp}=d_{OF}\theta_E/V_{\perp}$, onde $r_E$ é o raio do anel de Einstein que depende da massa da lente  e $V_{\perp}$ é a componente da velocidade perpendicular a linha de visada observador-fonte. A variação do ângulo,  implica que a quantidade definindo a magnitude $x(t)=\beta(t)/\theta_E$ (ver seção 3.2 e 3.3) é uma função do tempo que depende da massa da lente através do anel de Einstein (ver Eq. \ref{Eq8}). 

Todas as quantidades desconhecidas, incluindo $V_{\perp}$ podem ser estimadas combinando medidas cinemáticas e óticas com a variação máxima da curva de luz.  
%o que permite estimar a massa da lente contida na expressão do anel de Einstein ($\theta_E$).  pois a  velocidade ortogonal pode ser obtida da velocidade angular da fonte. 
%O tempo$\delta t$ no qual a fonte $x=x_{min}$ at´$x=1$ é dado pelo máximo da magnificação.
O método da curva de luz pode ser aplicado a 2 estrelas ou um par planeta-estrela, onde uma das estrelas ou planeta é a lente em movimento. 

%O problema é que  fração de estrelas ou planetas que oderiam estar dentro do raio de pEinstein dos MACHOs e EROS\footnote{A colaboração OGLE concentrou-se no Bojo da Via Láctea.} é muito pequena. A identificação dos eventos não é trivial. A cada momento, apenas uma estrela num  milhão sofre eventos de microlentes na galáxia. Assim,  um monitoramento de milhões de estrelas é necessário. 

Em linhas gerais, a conclusão final dos projetos MACHOs e  EROS com relação a possibilidade de descrição da matéria escura revelou-se negativa: \textit{Objetos compactos com massas $m\leq 30 M_{\odot}$ não são suficientemente abundantes para formar halos escuros, ou seja, não explicam as curvas de rotação.} 

Sua  abundância cósmica  inferida do halo da Via Láctea seria inferior a 20-25\%. Uma análise posterior (2003) obteve uma contribuição dos MACHOs para a matéria escura $< 5\%$ \cite{S2003}.

\textit{Se os objetos compactos -  MACHOS - não estavam na periferia da galáxia onde estariam?} Talvez  no Bojo, uma região de maior densidade estelar. Na verdade, o projeto OGLE concentrou-se  na região mais interna da Via Láctea, onde os efeitos de microlentes seriam mais abundantes. Os dados analisados a partir de 2003, mostrariam que o OGLE teria mais sucesso na busca dos eventos de microlentes. Em particular, na busca de exoplanetas; uma possibilidade sugerida por Mao e Paczynski \cite{BP1996,MP1991}.  

Os desenvolvimentos técnicos  posteriores  do OGLE e dos novos projetos na área como, MOA\footnote{MOA\,$\equiv$ \,Microlensing\, Observations\, in \,Astrophysics}, MICROFun\footnote{MICROFun\, $\equiv$ \,Microlensing\, Follow-Up\, Network}  e  PLANET\footnote{ PLANET\,$\equiv$ \, Probing\, Lens\,  Anomaly\,\, NETwork},  permitiriam a descoberta de exoplanetas a partir de 2004. 

O método é relativamente simples. Quando observamos a curva de luz de uma fonte distante cuja lente em movimento relativo é uma estrela circulada por um (ou mais planetas),  pequenas anomalias (``lombadas") podem ser observadas na curva de luz. 

\begin{figure}
\centering
\includegraphics[width=3.3truein,height=2.6truein]{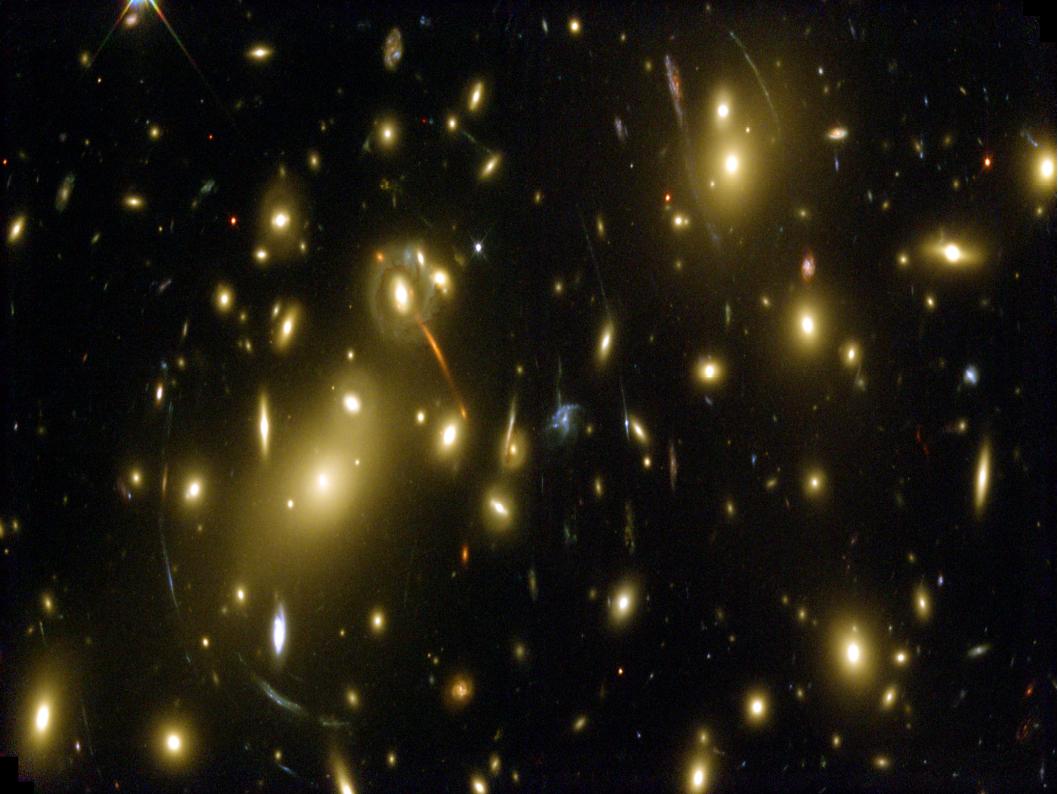}
\caption{Imagens múltiplas, arcos gigantes e pequenos arcos. Todas as imagens são vistas no grande aglomerado Abell 2218, no redshift $z=0,175$, um dos mais espetaculares sistemas de arcos variados (Crédito da imagem: Space Telescope Science Institute).}\label{Fig14}
\end{figure}

Na \textbf{Figura \ref{Fig13}} mostramos um exemplo concreto, imagiado  pela colaboração OGLE em 2006\footnote{O primeiro evento planetário de microlentes foi obtido em 2004, com massa $M_p=1,5\,M_{J}$  e um raio orbital de $\simeq$ 3 UA \, \,\cite{B2004}.}.  
% P
% p
%A curva de luz observada da microlente OGLE-2005-BLG-390
%modelo de evento e melhor ajuste plotado em função do tempo. 
O conjunto de dados em tempo real consiste de 650 pontos do PLANET Danish (ESO La Silla, ponto  vermelhos), PLANET Perth (azul), PLANET Canopus (Hobart, ciano), RoboNet Faulkes North (Havaí, verde), OGLE (Las Campanas, preto) e MOA (Observatório do Monte John, castanho). No canto superior direito vemos o \textit{"zoom"} da anomalia planetária, cobrindo um intervalo de tempo de 1,5 dias. A curva sólida é o melhor modelo de lente binária incluindo um planeta de massa, $M_p \simeq 5,5\,M_{terra}$, localizado a uma distância  $\simeq 2,6 UA$ da estrela. 

Atualmente, mais de 60 exoplanetas já foram descobertos por esta técnica, nascida dos contributos seminais de Paczynski e colaboradores (1986-1991).

\subsection{Lentes Fortes e Fracas}

Aglomerados de galáxias são as maiores estruturas do Universo   gravitacionalmente ligadas que podem estar em equilíbrio hidrostático (virializadas). Aglomerados ricos chegam a ter mais de $10^{3}$ galáxias, massas no intervalo $10^{14} M_{\odot} \leq M \leq 10^{16}M_{\odot}$ e diâmetros entre 1-2 Mpc. A composição inclui a massa estelar nas galáxias, um gás quente (plasma) no meio intergalático e matéria escura. Tal como as galáxias, os aglomerados deformam o espaço-tempo nas partes internas e na vizinhança, encurvando a luz das fontes distantes. Nas regiões internas, a densidade é alta e pode defletir os feixes da frente de onda,  produzindo imagens múltiplas.   

Na \textbf{Figura \ref{Fig14}} (ver também  \textbf{7b}), vemos que as galáxias  multiplicam o efeito de lentes formando extraordinários arcos, cuja análise espectroscópica revela o mesmo redshift e, portanto, são lentes de um mesmo objeto distante. Os arcos gigantes aparecem  curvados  na direção da região mais densa, o centro do aglomerado.  Além disso, vemos da figura que sua extensão  tangencial (relativa ao centro do aglomerado) é bem maior  do que a largura radial. A formação de arcos gigantes e pequenos, descoberto nos anos 90 (ver também \textbf{seção 3.5}),   caracteriza o regime de lentes fortes em aglomerados.    
%Os arcos  gigantes  funcionam com sondas da região central. 

No limite que as deflexões causam pequenas modificações nas propriedades do objeto (brilho, posição, tamanho  e forma), mas não fenômenos visualmente impressionantes, como múltiplas imagens ou arcos, o processo é chamado de lente fraca.  Em geral são induzidas pelas flutuações na densidade média do Universo. As distorções observadas na periferia dos aglomerados são uma combinação homogênea de forma  (convergência) com o cisalhamento (``shear"), provocados pelo efeito de lente fraca ao longo da linha de visada \cite{NB1996}.
 
 \begin{figure}
\centering
\includegraphics[width=3.2truein,height=2.6truein]{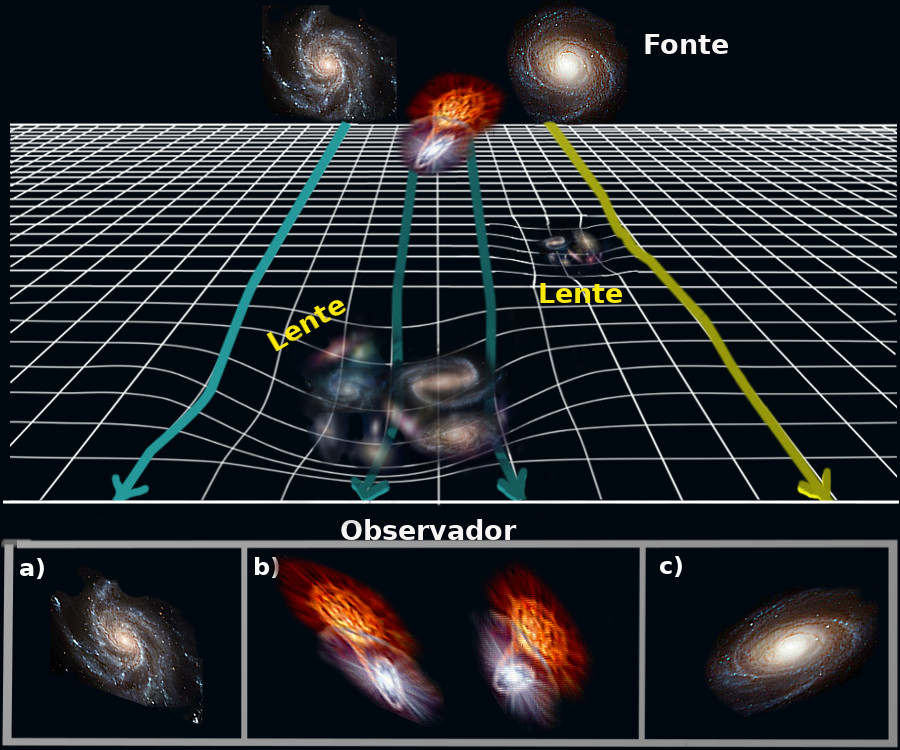}
\caption{No caso dos exemplos (a) e (c) temos imagens distorcidas devido ao efeito de lente fraca, enquanto em (b) temos a formação de múltiplas imagens devido ao efeito de lente forte.} \label{Fig15}
\end{figure}
 
 Na \textbf{Figura \ref{Fig15}}, mostramos o efeito qualitativo de lentes forte e fraca. No primeiro caso a luz passa mais distante do núcleo do aglomerado e o efeito  pode ser uma mera distorção, um  cisalhamento (``shear") na imagem; um círculo aparece como uma elipse [inserções (a) e (c)]. No caso de lentes fortes, a luz das fontes lenteadas passam na parte mais interna dos aglomerados sofrendo distorção,  amplificação e multiplicação de imagens [inserção (b)].  
 
 Lente gravitacional por aglomerados de galáxias é uma das principais ferramentas cosmológicas para acessar a matéria escura e o Universo distante. Assinaturas de lentes fortes funcionam como  sondas da distribuição da matéria mais interna dos aglomerados de galáxias, permitindo a localização do núcleo dos clusteres e a determinação de sua massa. A ampliação de fluxo aparente transforma aglomerados de galáxias em telescópios gravitacionais, que podem ser usados para estudar galáxias em  altos redshifts que seriam, caso contrário, muito fracas para serem observadas. 
 
 %a abundância de galáxias de fundo e outras fontes fortemente lenteadas, aparecendo como arcos gravitacionais e imagens múltiplas, pode ser comparada com as %previsões da eficiência das lentes de halos na escala de cluster e também nas 
 
 O estado dinâmico do cluster e a presença de substruturas podem também ser investigados via lentes fracas. Medidas de massa com lentes fracas e de raios-X, e a velocidade de dispersão podem ser comparadas  com simulações testando  a estrutura da ``teia  cósmica" e também saber se o aglomerado está em equilíbrio ou se as diferentes componentes estão interagindo \cite{Oliveira2017,CL2004,Meneghetti2013,Dodelson2017}.   
% P
% p
\section{Idade Contemporânea (2006-2019)}

Em 1933, Zwicky já havia inferido  a existência de matéria escura, estimando a dispersão de velocidade das galáxias na região central dos aglomerados \cite{Z33}. Na idade moderna sua conjectura foi adotada devido a dois efeitos: (i) as curvas de rotação (seção 4.1) e, (ii) existência dos arcos como um efeito de lentes nos aglomerados. 

\begin{figure}
\centering
\includegraphics[width=3.2truein,height=2.6truein]{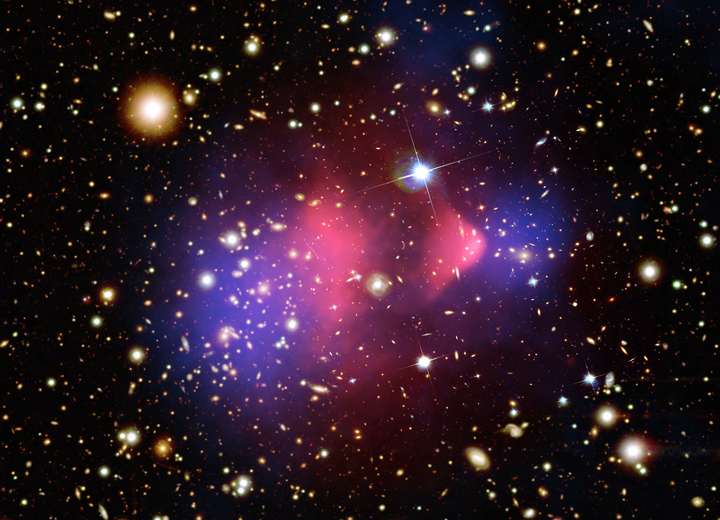}
\caption{Aglomerado de galáxias $1E 0657-56$. Devido ao  perfil ficou popularmente conhecido como aglomerado bala (\textit{bullet Cluster}). O menor cluster à direita atravessa o maior à esquerda. Nessa imagem, dados de raios-X (parte avermelhada) e o mapeamento da massa obtido de lentes fracas (azul) são sobrepostos a imagem óptica do HST (detalhes no texto). Crédito: \textit{Raios-X}: NASA/CXC/CfA/; \textit{Mapa de Lentes}: NASA/STScI; ESO WFI; Magellan/U.Arizona/; \textit{ótico}: NASA/STScI; Magellan/U.Arizona.} \label{Fig16}
\end{figure}

Mais recentemente (2006), a existência da matéria escura foi também sugerida pela  primeira fusão observada de dois aglomerados; um evento que ficaria conhecido como aglomerado bala (\textit{"Bullet Cluster"}). As LGs  desempenharam um papel crucial nessa descoberta (\textbf{Figura \ref{Fig16}}),   considerada até o presente a evidência mais direta da matéria escura \cite{Clowe,Oliveira2018}.   
% P
% p
\subsection {O Aglomerado Bala}
Um caso especial de mapeamento do perfil de densidade e massa total por lentes ocorre na fusão de 2 aglomerados de galáxias. O primeiro e mais popular exemplo, é o aglomerado bala, 1E 0657-56, considerado o evento mais energético conhecido no Universo desde o Big Bang\footnote{Uma simulação mostrando a formação do \textit{Bullet Cluster} pode ser vista no site https://www.youtube.com/watch?v=eC5LwjsgI4I}. Seu mapeamento é uma combinação de duas técnicas: raios-X e lentes gravitacionais.

Na  \textbf{Figura \ref{Fig16}}, vemos o resultado simulado do processo de fusão dos dois aglomerados de galáxias.
% P
% 
%\footnote{Crédito da imagem em raios-X: NASA/CXC/CfA/M.Markevitch et al.; Optical: NASA/STScI; Magellan/U.Arizona/D.Clowe et al.; Lensing Map: NASA/STScI; ESO WFI; Magellan/U.Arizona/D.Clowe et al. Retirado do site: http://chandra.harvard.edu/photo/2006/1e0657/, em 23/04/2019. } \textit{(Bullet Cluster)},
O plasma quente (avermelhado) foi detectado pelo observatório Chandra\footnote{NASA X-Ray Observatory} e contém a maior parte da matéria normal ou bariônica dos dois aglomerados \cite{Markevitch}. O perfil em forma de bala à direita é o gás do aglomerado menor e mais veloz, atravessando o gás quente do maior aglomerado durante a colisão. As áreas azuis  assinalam onde as lentes gravitacionais encontraram a maior parte da massa dos aglomerados (note que  o centro de gravidade não coincide com o da massa bariônica). Devido a interação, o movimento do gás  defasou-se em relação ao da matéria escura, uma componente que interage apenas gravitacionalmente.

Dizendo de outra forma, o plasma de cada cluster foi retardado pela força média (mútua), de origem viscosa, semelhante à resistência do líquido quando  atravessado por uma bala. Durante a colisão,  a matéria escura (não interagente) dos dois aglomerados, moveu-se à frente do gás quente, produzindo a separação do material escuro e bariônico vista na imagem. O resultado mostra que a matéria escura é necessária na escala dos aglomerados e reafirma seu status de componente material (não relativística) dominante no Universo\footnote{A existência do evento \textit{bullet cluster} contradiz as teorias de gravidade que alteram a lei de força newtoniana \cite{M1983}, coletivamente denominadas do tipo MOND (\textit{Modified Newtonian Dynamics}).}\,\cite{MOND2001}.

\subsection{Universo Acelerado e Lentes Gravitacionais}

A descoberta da aceleração do Universo em 1998, através das observações de Supernovas do tipo Ia, estabeleceu um novo paradigma na Cosmologia \cite{Riess1998,Perlmutter1999}. Em 2006, o modelo acelerado,  plano,   composto  por bárions ($\sim 5\%$), matéria escura ($\sim 25\%$)\, e energia escura ($\sim 70\%$), já era considerado o modelo cosmológico padrão. Mas não era  conhecida ainda a natureza da energia escura e da matéria escura, o chamado setor cosmológico escuro. 

Em 2006, o projeto Dark Energy Task Force (DETF) foi criado com o objetivo de determinar a natureza da Energia Escura \cite{DETF2006}. A questão era decidir se o agente acelerando o Universo era a constante cosmológica (\textbf{$\Lambda$}), associada a densidade de energia do vácuo,  ou algum outro tipo de campo. Várias abordagens são possíves. A metodologia adotada  foi a combinação de 4 diferentes subáreas (i) Oscilações Acústicas dos Bárions, (ii) Aglomerados de Galáxias, (iii) Supernovas, e (iv) Lentes Gravitacionais. Algumas abordagens recentes e resultados envolvendo a área de lentes gravitacionais serão descritas a seguir.

\subsection{SNe Ia e Lentes Fracas}

\begin{figure}
\centering
\includegraphics[width=3.3truein,height=3.0truein]{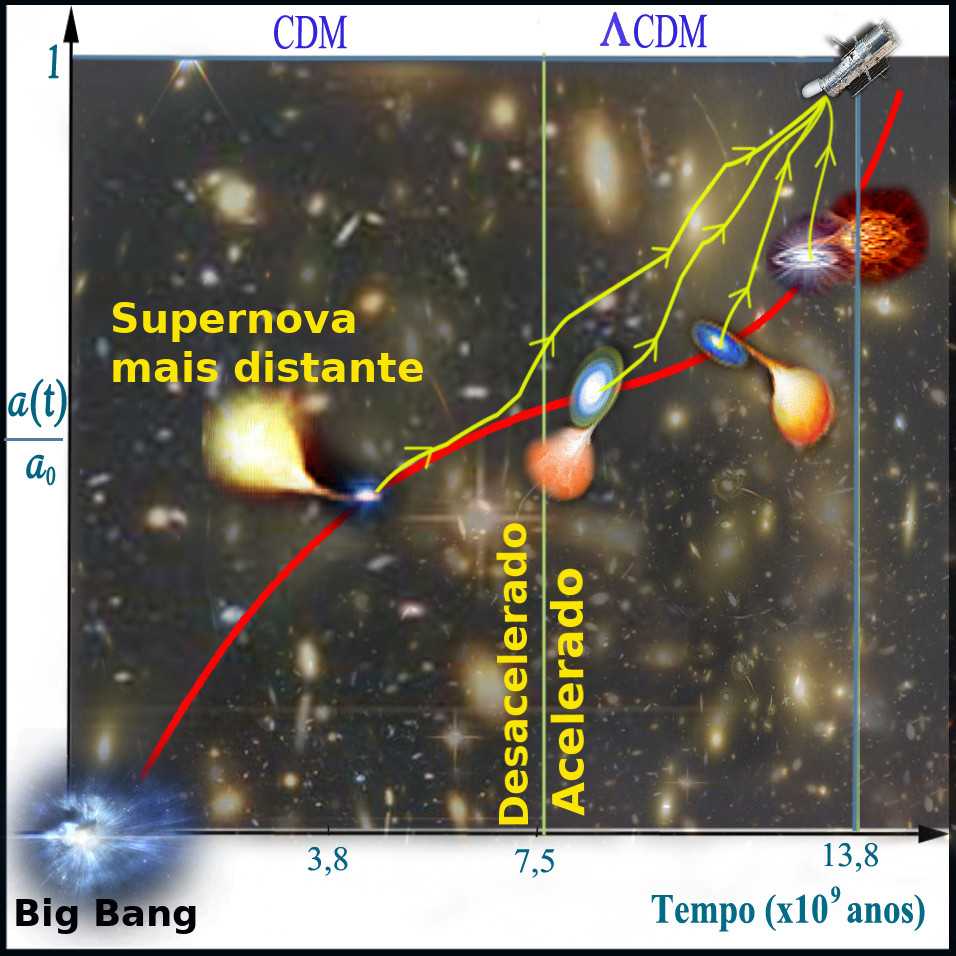}
\caption{Supernovas em diferentes \textit{redshifts} sofrendo o efeito de lente fraca ao longo do percurso. A primeira linha vertical separa a fase de expansão desacelerada (CDM) da fase acelerada ($\Lambda$CDM). Já a segunda linha vertical indica o tempo presente. Note que a supernova mais distante foi imagiada quando o universo ainda estava na fase desacelerada.} \label{Fig17}
\end{figure}

As SNe Ia são também lenteáveis pela estrutura do Universo, magnificando (ou demagnificando) seu brilho, dependendo se a linha de visada para o SNe Ia passa por regiões de sobredensidades (subdensidades) de matéria. Com suficiente número de objetos e uma modelagem adequada das lentes fracas \cite{Wang05,R2017}, os dados poderiam ser  utilizados para estimar parâmetros cosmológicos usando a distribuição de SNe Ia (\textbf{Figura \ref{Fig17}}) cruzada com a densidade observada ao longo da linha de visada, ou mesmo usando sua amplificação (à "la Paczynsky") para avaliar as propriedades do halo da matéria escura dos aglomerados  \cite{Goliath00,Jonsson08}. 

Em 2014, a magnificação das SNe Ia foi  marginalmente detectada por Smith \textit{et al.} \cite{Smith14}. Os autores  correlacionaram as
magnitudes de SN Ia da amostra SDSS\footnote{SDSS $\equiv$ Sloan Digital Sky Survey} com a densidade observada ao longo da linha de visada. No entanto, a distribuição é sensível à estrutura de grande e pequena escala que é bem mais  difícil de modelar. 

 Os dados para  $z>1$, ou seja, quando o universo ainda estava desacelerado, podem ser obtidos medindo-se os redshifts da galáxia hospedeira. A princípio, tal técnica poderá gerar um conjunto de dados de SN Ia, proporcionando uma interessante sonda cosmológica no intervalo  crítico de redshifts $1 < z <2$.
 % P
%p 
\subsection{SNe Ia e Lentes Fortes}

 Refsdal mostrou que a cosmografia com \textit{``time delay}", envolve a diferença no tempo, $\Delta \tau$, de chegada da luz nas imagens múltiplas \cite{Re1964b}.  Combinando as medidas com um modelo para o potencial da lente, pode-se mostrar que o atraso depende de uma quantidade envolvendo as distâncias cosmológicas da fonte e da lente. Resulta que esta quantidade é inversamente proporcional a $H_{0}$ e fracamente sensível aos outros parâmetros cosmológicos, de modo que  $\Delta \tau \propto H^{-1}_{0}$. Portanto,  medidas de time delay permitem inferir o valor da constante de Hubble. Alguns autores chamam esse método de TDSL (\textit{time delay strong lensing}).

Estimativas de $H_0$ via TDSL são independente dos métodos locais, baseados na escada de distância cósmica (Cefeídas, Supernovas, etc.) ou em altos redshifts, via radiação cósmica de fundo, como as obtidas através dos satélites WMAP e Planck. Inicialmente, o TDSL fornecia valores de $H_0$ bem mais  baixos do que os baseados nas Cefeídas. No entanto, o método foi empregado recentemente com relativo sucesso utilizando imagens múltiplas de quasares \cite{Treu2016,Grillo2018}. Fontes gravitacionalmente lenteadas com múltiplas imagens, provenientes de Supernovas (ou de quasares),  podem fornecer uma valiosa contribuição para as medidas de $H_0$.

% P
% p

\begin{figure}
\centering
\includegraphics[width=3.3truein,height=2.9truein]{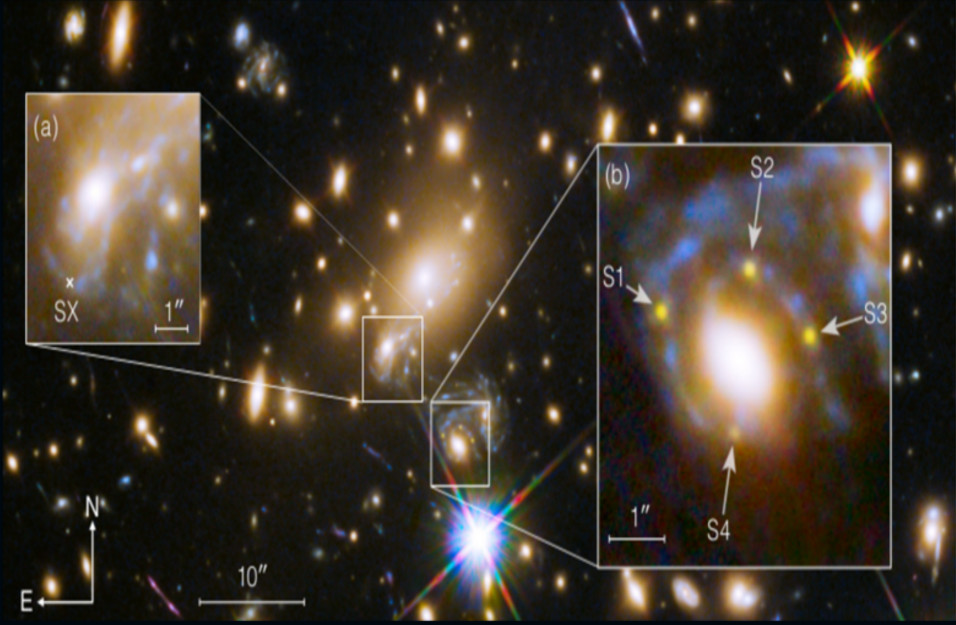}
\caption{Descoberta da supernova "Refsdal" \,no aglomerado de galáxias, MACS $J1149.6 + 2223$. A galáxia hospedeira tem \textit{redshift} $z = 1,49$. Nos paineis vemos: (a) localização da quinta imagem detectada, observada pela primeira vez em dezembro de 2015, (b)  primeiras quatro imagens, S1-S4, observadas quando a SN ``Refsdal"\, foi detectada em 2014 \cite{Kelly2015}.  Disponível no artigo \cite{Treu2015} (Crédito: NASA, ESA/Hubble).
} \label{Fig18}
\end{figure}

Na \textbf{Figura \ref{Fig18}}, mostramos a chamada Supernova de\\``Refsdal"\, e suas várias imagens. O \textit{redshift} da SN é o mesmo da  galáxia hospedeira\, ($z=1,49$). Em destaque as quatro imagens (S1-S4) formando uma "Cruz de Einstein" [inserção (b)]. Na inserção (a) vemos a quinta imagem observada em 2015.

O método TDSL está sendo aplicado  por pesquisadores envolvidos na colaboração COSMOGRAIL\footnote{COSMOGRAIL\,$\,\equiv$ Cosmological\, Monitoring of Gravitational \,Lensing} e  HOLiCOW\footnote{HOLiCOW $\equiv$ $H_0$ Lensing in COSMOGRAIL’s Wellspring}. Neste ano, eles  obtiveram  um valor  de $H_0$, relativamente alto e preciso,  utilizando o TDSL com imagens de quasares. Um resultado que nos leva a discutir  a tensão Supernova versus RCF, referente as medidas de $H_0$ e sua conexão com lentes gravitacionais. 
% P
% p
\subsection{Tensão no Valor de Ho e TDSL}

$\Lambda$CDM plano é  considerado  o atual modelo cosmológico padrão. No entanto, desde que  os resultados de diferentes experimentos cosmológicos cresceram em precisão, começaram a surgir alguns pontos de tensão. O mais notável e desafiador deles, é a discrepância entre as medições da constante de Hubble, a partir de análises  da radiação cósmica de fundo (RCF) -  pelas missões dos satélites WMAP e Planck - comparadas com as medidas locais,  baseadas em Cefeídas e Supernovas, obtidas através do método padrão (diagrama de Hubble-Sandage). 

Por um lado, as medidas de $H_0$ baseadas na RCF são oriundas de altos \textit{redshifts}, atribuem um valor menor e são dependentes do modelo padrão, e do outro, os valores  inferidos com base na escada de distância cósmica, embora dependentes da calibração, são locais, ou seja, obtidas em \textit{redshifts} extremamente baixos. Neste caso, os valores de $H_0$ são maiores  e independentes de modelo cosmológico. 
% P
% p
A discrepância dos valores de $H_0$ obtida pelos dois  métodos vem crescendo desde 2011. Paralelamente, a precisão nas medidas de $H_0$ por cada método também aumentou, tornando mais acentuada a divergência. Entre 2011 e 2014, as  medidas locais chegaram a uma precisão melhor do que $3\%$.  Riess et al. \cite{Riess2011} obtiveram $H_0 = 73,8 \pm 2,4$ km.$s^{-1}.Mpc^{-1}$ usando Cefeídas e Supernovas, enquanto  Lima e Cunha \cite{LC2014} com a mesma precisão, obtiveram o valor  $H_0 = 74,1 \pm 2,2 $ km.$s^{-1}.Mpc^{-1}$. Esse último resultado emergiu de uma análise  combinando diferentes observações em \textit{redshifts}\, intermediários (z $\simeq$ 1); e tal como o Planck também depende do modelo $\Lambda$CDM. 
% P
% p

Na mesma época (2013),  as colaborações do WMAP e Planck obtiveram,  $H_0 = 70,0 \pm 2,2 $ km.$s^{-1}.Mpc^{-1}$ e $H_0 = 67,4 \pm 1,4 $ km.$s^{-1}.Mpc^{-1}$, respectivamente, com incertezas de $\sim$ 2\%. Note que dentro das incertezas das medidas, os resultados da RCF são consistentes entre si, mas discrepantes com as medidas mais locais; sendo estatisticamente mais significante no caso do Planck.  Nessa altura estava  aceso o sinal vermelho entre astrônomos e cosmólogos.\,\textit{A discrepância em $H_0$ iria estacionar, aumentar ou diminuir?}

\begin{figure}
\centering
\includegraphics[width=3.3truein,height=2.9truein]{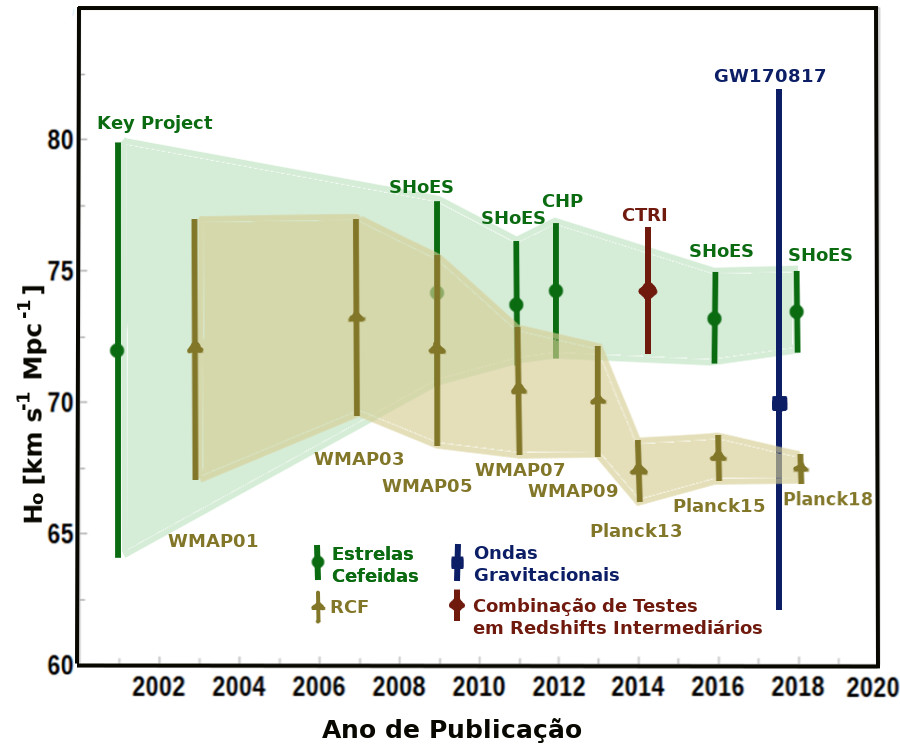}
\caption{Evolução da Tensão em $H_0$. Pontos e região  sombreada de verde foram obtidos com Cefeídas e Supernovas, enquanto os da região creme são determinados pelas medidas da RCF. O ponto em vermelho está relacionado a uma combinação de testes em redshifts intermediários (CTRI, Lima \& Cunha 2014 \cite{LC2014}). A barra  azul é uma medida com ondas gravitacionais (adaptado de Freedman  \cite{Freedman2017}).} \label{Fig19}
\end{figure}

Na \textbf{Figura \ref{Fig19}}, mostramos a evolução das medidas de $H_0$  e o aumento contínuo da discrepância entre os 2 métodos. A discrepância aumentou entre 2015 e 2018, a medida que a precisão aumentava  \cite{planck2015,Riess2016,planck2018}. Em 2019,  utilizando dados de 70 Cefeídas imagiadas na grande nuvem de Magalhães, a colaboração SH0ES\footnote{SHOES  $\equiv$ Supernovae, $H_0$, for the Equation of State of Dark Energy} obteve $H_0=74,03\pm 1,42 $km.$\,s^{-1}.Mpc^{-1}$; valor a ser  comparado com   $H_0=67,4\pm\,0,5\,$km.$\,s^{-1}.Mpc^{-1}$, publicado pelo Planck (2018). Tais resultados implicam  que a diferença entre os valores de $H_0$ 
medido localmente e pelo Planck, $\Delta H_0 = 6,6 \pm 1,5$ km.$\,s^{-1}.Mpc^{-1}$, bem além de uma discrepância meramente aleatória (4.4$\sigma$) \cite{riess2019}. A discrepância parece consolidada, mas sua explicação continua desconhecida.

% P, % p  
 Qual será o valor correto de $H_0$, o de altos ou de baixos redshifts? Mais importante no presente contexto: \textit{Qual o papel das lentes gravitacionais nessa crise?} 
 
 Cosmografia de ``\textit{time delay}"\, (TDSL) usa atrasos de tempo de lente gravitacional e pode  medir a constante de Hubble. Como observado no final da seção anterior, a idéia vem sendo implementada no âmbito do projeto H0LiCOW \cite{Suyu2017} em colaboração com o  COSMOGRAIL \cite{Courbin2018}. 
 Como resultado, supondo um modelo $\Lambda$CDM plano, obtiveram para o QSO 1206 + 4332 com duas imagens, o seguinte valor: $H_0=72,5^{+2,1}_{-2,3}\,$km.$\,s^{-1}.Mpc^{-1}$ com precisão de  3\%\,. Uma  medida independente da escada de distância cósmica e de outros testes cosmológicos, mas dependente do modelo $\Lambda$CDM plano \cite{Birrer2019}. 

Não sabemos se o resultado acima resolve a tensão sobre o valor correto de $H_0$. Talvez um solução definita venha com o telescópio LSST\footnote{LSST\,$\equiv$ \, Large Synoptic Survey Telescope.} que está sendo construído no Chile. Dotado de um  espelho de 8,4 metros e projetado para entrar em operação em 2022, o LSST  será capaz de mapear todo o céu visível. Com o LSST, espera-se obter milhares de transientes com lentes fortes (supernovas e quasares) que deverão turbinar e aumentar a precisão da cosmografia do atraso de tempo.

\section{Lentes e Ondas Gravitacionais: O Futuro?}
% P
% p
Recentemente, as colaborações LIGO\footnote{LIGO $\equiv$ Laser\, Interferometer \, Gravitational-Wave Observatory} e VIRGO\footnote{VIRGO $\equiv$ Interferômetro Europeu de Ondas Gravitacionais} detectaram  ondas gravitacionais (OGs) emitidas durante a fusão de pares de buracos negros e choque entre estrelas de nêutrons \cite{Abbott2016a,Abbott2016b, Abbott2017}. Esta bem sucedida previsão da relatividade geral, pressiona por esforços observacionais envolvendo a próxima geração de detectores de OGs no solo e no espaço, planejados para "ouvir"\, bandas de freqüência emitidas por uma grande variedade de fontes. Numa astronomia multimensageira, é natural que a interface unindo lentes e OGs seja também investigada. 

 %Atualmente, estamos ainda caracterizando as OGs e suas fontes independentemente de sua abundância cósmica ou se estão muito distantes \cite{Abbott2017,Maggiore2008} ou próximas do sistema solar. Fontes mais próximas e menos extremas também tem sido investigadas, pois a pequena distância  pode compensar sua intensidade relativamente baixa, fornecendo  alvos  que permitem uma observação sem interrupções \cite{CSL2018,Tamanini2019}. 
 
 Na relatividade geral, grávitons  são partículas sem massa e, tal como os fótons, seguem  geodésicas nulas. Isto significa que o campo gravitacional  deverá também agir como um meio refrator sobre as OGs (ver \textbf{Apêndice A}). Como acontece com  feixes de luz, discutidos na aproximação da ótica geométrica\footnote{Ver \textbf{rodapé 9}.}, os feixes de grávitons serão também  defletidos.  

No entanto, a escala macroscópica característica das lentes gravitacionais é o raio gravitacional ($L=R_g=2GM/c^{2}$). Isto significa que o limite da ótica geométrica gravitacional  ($\lambda_g <\,<  R_g$), pode  não ser satisfeito em muitas situações.  Por exemplo, no caso do LIGO ($10km <\lambda_g < 10^{4}km$), a ótica geométrica será apropriada quando a lente for uma galáxia, mas não um objeto ultracompacto, pois neste caso  $\lambda_g \gtrsim  R_g$. Portanto, efeitos ondulatórios como a difração, em geral,  não poderão ser ignorados, tornando a fenomenologia do lenteamento  das OGs bem mais rica e complexa. 

Ondas  gravitacionais são  perturbações no espaço-tempo,  potencialmente, afetando  o campo da lente e suas imagens. No ótico, a perturbação da onda pode afetar as "cintilações"  alterando o \textit{"time-delay"} entre a fonte e as imagens. Para o quasar duplo 0957 + 561 (A,B), o atraso medido é de 415 dias. Isto significa que possíveis contribuições de ondas gravitacionais podem estar incluídas no processo e precisam ser subtraídas. 

Por outro lado, se imagens gravitacionais múltiplas forem  identificadas, outras possiblidades serão abertas. Por exemplo, parâmetros cosmológicos medidos via OGs com o efeito de ``sirenes cósmicas" \cite{L2017,F2019}, fornecerão uma estratégia alternativa de solução para a presente tensão sobre a constante de Hubble. O caso mais óbvio podemos denominar  GTDSL (\textit{Gravitational Time Delay Strong Lensing}), em  perfeita analogia com o TDSL. 

Similarmente, sistemas exoplanetários de período ultracurtos são também fontes interessantes de ondas gravitacionais para a próxima geração de detectores, pois estão dentro da curva de sensibilidade  do LISA\footnote{LISA $\equiv$ Laser Interferometer Space Antenna} e outros instrumentos planejados. Reciprocamente, as ondas gravitacionais emitidas  podem fornecer uma janela complementar para identificar planetas extrasolares via microlentes gravitacionais, um novo procedimento que pode ser ainda mais eficiente do que os métodos disponíveis (como o tempo de trânsito) baseados em instrumentos e técnicas ópticas \cite{CSL2018,Tamanini2019}.
% P
% p
\section{Comentários Finais \& Perspectivas}

Um século depois da legendária observação do eclipse solar total,  confirmando o resultado da deflexão  relativística prevista por  Einstein, podemos  afirmar que as lentes gravitacionais constituem uma  área de pesquisa consolidada da Astronomia. 

Na verdade, no final dos anos 80,  a lista de descobertas científicas em lentes gravitacionais ainda era modesta, embora os progressos realizados até meados da idade moderna já fossem animadores. Em outras palavras,  apesar dos inúmeros obstáculos, valia a pena investigar as aplicações de lentes gravitacionais; uma área oferecendo novas técnicas na abordagem de problemas fundamentais \cite{ELTurner}. De fato, além de proporcionar  vários  ``efeitos especiais"\, como os anéis e a cruz de Einstein, a magnificação variável da curva de luz de estrelas fontes e a formação de arcos gigantes em aglomerados, revelou-se também  uma disciplina com ramificações em várias direções; desde a busca  dos objetos escuros até as mais desafiadoras questões cosmológicas. 

\begin{figure}
\centering
\includegraphics[width=3.4truein,height=2.9truein]{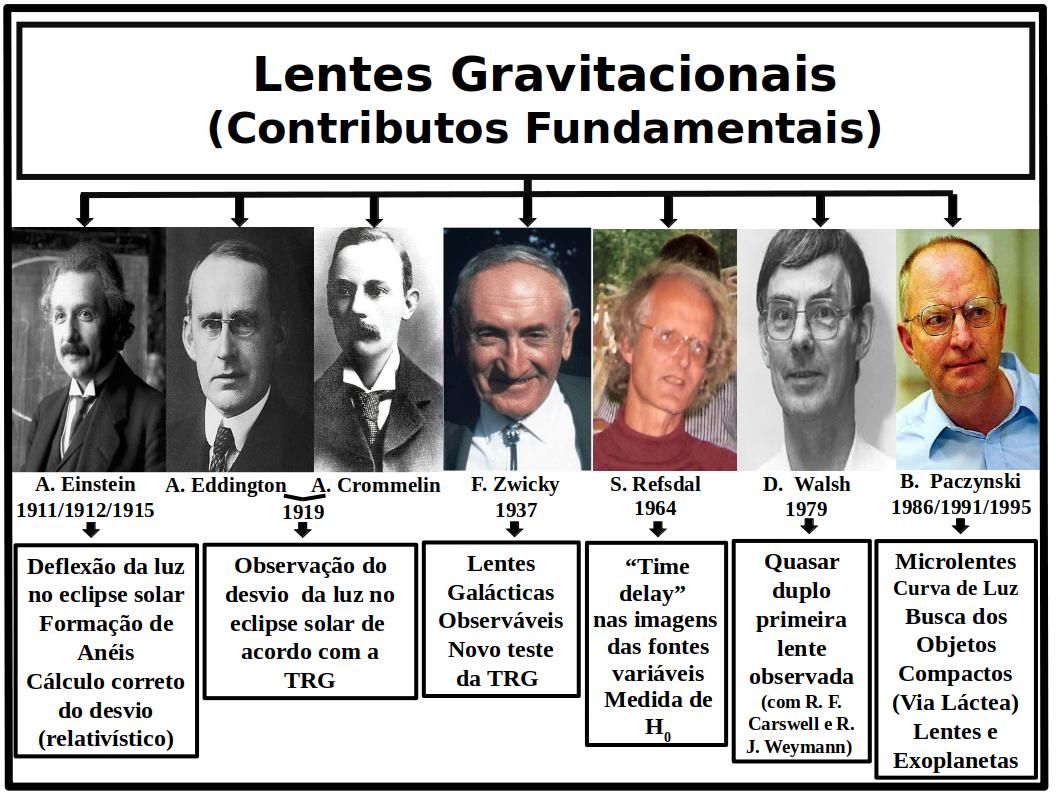}
\caption{Fundadores da área de lentes gravitacionais (Idade Antiga e Moderna). O período entre 1938-1962, foi a idade média das lentes gravitacionais. Seu renascimento ocorreu nos anos 60-70 com os trabalhos teóricos de Refsdal e a descoberta do quasar duplo 0957 + 561 A\, \&\, B (infelizmente, os autores não encontraram fotos de Carswell e Weymann). Tais resultados foram consolidados  pelas diferentes aplicações da curva de luz de Paczinski, na sua mais criativa década.}\label{Fig20}
\end{figure}

Na \textbf{Figura \ref{Fig20}}  mostramos os pioneiros da teoria de lentes gravitacionais das eras antiga e moderna (cf. \textbf{Figuras \ref{Fig1} e \ref{Fig8}}). Passadas as fases mais heróicas, digamos, até meados dos anos 80, estamos agora vivenciando a era da precisão observacional.  Entramos numa fase de projetos e colaborações de grande envergadura, tanto em recursos humanos e financeiros quanto em sofisticação tecnológica. Isto significa que os objetivos científicos são bem definidos, mas a coleta e o tratamento de dados revela-se uma tarefa colossal, tornando todo e qualquer avanço um trabalho de caráter bem mais coletivo. Por isso evitamos uma cronologia pontuando nomes na fase contemporânea. 

Naturalmente, a nova realidade  não exclui o exercício individual da criatividade, nem tão pouco a possibilidade de uma quebra de paradigma na disciplina. Nesse sentido, uma atenção especial deverá ser dedicada as novidades  oriundas  dos campos emergentes envolvendo  lentes, desde a busca por exoplanetas até as ondas gravitacionais. 

Descobrir novos planetas é um dos grandes objetivos da astronomia moderna com consequências para o estudo da origem e preservação da vida em diferentes ambientes do Universo. A busca por planetas extrasolares na chamada zona habitável\footnote{Um planeta está na zona habitável quando sua distância até a estrela hospedeira é compatível com a existência de água líquida.} tem sido reforçada por satélites exclusivamente dedicados a tarefa. Nessa conexão, microlentes hoje é parte do conjunto de técnicas utilizadas nas descobertas, tais como: imagiamento direto, tempo de trânsito e as medidas de velocidade radial. Mais importante ainda, microlentes  explora regimes de massas e tamanhos de órbitas diferentes das outras técnicas (\textbf{seção 4.1}). 

Sem dúvida, podemos afirmar que a área de lentes gravitacionais  tem tido um sucesso espetacular no estudo de fenômenos astronômicos; varrendo desde objetos individuais, como planetas, estrelas, galáxias e aglomerados, até a distribuição de massa em várias escalas; incluindo ainda medidas de vários parâmetros cosmológicos e uma possível solução da tensão envolvendo as medidas da constante de Hubble em altos e baixos redshits. 

Todas essas realizações garantem expectativas extremamente positivas para os próximos vinte anos. Nesse contexto, não será surpreendente se a detecção futura do lenteamento das ondas gravitacionais abrir uma nova janela de oportunidades para nossa compreensão do Cosmos.

\vskip 0.3cm
\textbf{Agradecimentos:} JASL agradece ao CNPq,  CAPES (PROCAD 2013) e   FAPESP (Projeto LLAMA).    
\vskip 0.3cm

\noindent \textbf{\large Apêndice A: Campo Gravitacional como um Meio Refrator}
\vskip 0.4cm

Neste Apêndice, mostramos que o campo gravitacional pode ser interpretado como um meio refrator de índice de refração $ n > 1$. Isso significa que na presença da gravitação a velocidade aparente da luz, $c'=c/n$, é menor do que  a velocidade da luz no vácuo (distante da fonte).

Na chamada aproxima\c{c}\~ao de campo fraco, a m\'{e}trica do espaço-tempo pode ser escrita como:

\begin{equation}
g_{\mu\nu}\equiv \eta_{\mu\nu} + h_{\mu\nu}\,,
 \label{eq:weak-field}
\end{equation}
onde $\eta_{\mu\nu} = diag(+1,-1,-1,-1)$ é a métrica de Minkowski e $|h_{\mu\nu}|<<1$. 
Nesta aproximação, a geometria na presença do corpo defletor toma a seguinte forma: 
%\cite{LL1975c}:
\begin{equation}
ds^2 = \left(1+\frac{2 \Phi}{c^2}\right)~c^2dt^2 - \left(1-
\frac{2\Phi}{c^2}\right)~ dl^2\;,\label{LE}
\end{equation}
onde $\Phi(x)$ é o potencial newtoniano ($|\Phi|/c^2<<1$) e $|dx|= dl$, o elemento de linha espacial.
 
%Na aproxima\c{c}\~{a}o da \'{o}tica geom\'{e}trica, a escala sobre a qual o campo gravitacional varia \'{e} muito maior que o comprimento de onda da luz defletida. 
Sabemos que na relatividade geral, os fótons seguem geodésicas nulas,  $ds^2=0$.  Resolvendo a equação acima para a velocidade efetiva, $dl/dt = c'$, obtemos:

\begin{equation} 
{c'}\equiv\frac{dl}{dt}=c\left(\frac{1+\frac{2\Phi}{c^2}}{1-\frac{2\Phi}{c^2}}\right)^{1/2} \simeq c(1 +  \frac{2\Phi}{c^2}).  \label{Eq20}
\end{equation}
Portanto, como o potencial gravitacional para um objeto compacto \'{e} uma quantidade negativa, segue que a velocidade aparente da luz \'{e} menor na presen\c{c}a do campo gravitacional. Similarmente, podemos caracterizar o efeito da
propaga\c{c}\~{a}o da luz atrav\'es do \'{\i}ndice de refra\c{c}\~{a}o efetivo definido por:

\begin{equation}
n(\mathbf{x}) = \frac{c}{c'(\mathbf{x})}= 1 - \frac{2
\Phi(\mathbf{x})}{c^2}\;. \label{Eq21}
\end{equation}
Na  aus\^{e}ncia de campo gravitacional [$\Phi(\mathbf{x}) =0$], temos o valor do v\'acuo ($n = 1$). Tal como ocorre  na \'{o}tica geom\'{e}trica usual, um valor de $n > 1$ implica que a luz viaja mais lentamente do que no vácuo e  podemos dizer que a gravitação  atua como um meio refrator. Naturalmente, tal efeito  n\~{a}o deve ser confundido com a refra\c{c}\~{a}o ordinária ocorrendo, por exemplo, na atmosfera terrestre; um fen\^{o}meno associado ao fato da atmosfera gasosa ser constitu\'{\i}da por camadas de densidade vari\'{a}vel.
%Para o Sol a correção na superfície  $\Delta n/n \simeq 10^{-6}$.

\vskip 0.3cm

\noindent \textbf{\large Apêndice B:  Magnificação e Variabilidade}
\vskip 0.4cm
Neste Apêndice, apresentamos os detalhes do cálculo da magnificação ${\mathcal M}$  utilizados na subseção 3.3  para o caso da lente pontual (ver \textbf{ Figura 5}). 

Da expressão ${\mathcal M_{\pm}}=\frac{\theta{\pm}}{\beta}\frac{d\theta_{\pm}}{d\beta}$, e considerando os valores de $\theta_{\pm}$, mostra-se   que a magnificação de cada imagem pode ser escrita como: 

\begin{equation}
{\mathcal M_{\pm}}=\frac{\theta_{\pm}}{2\beta}
\left(\frac{\beta}{\sqrt{\beta^{2} + 4\theta^{2}_E}\pm 1} \right). \label{Eq22}
\end{equation}
onde $\theta_{\pm}$ são as soluções das imagens na equação (\ref{Eq10}). Introduzindo a quantidade normalizada,  $x=\beta/\theta_E$, temos:
\begin{equation}
{\mathcal M_{\pm}}=\frac{x^{2} + 2}{2x\sqrt{x^{2} +4}} \pm \frac{1}{2}. \label{Eq23}
\end{equation}
Finalmente, adicionando os dois valores, a magnificação total é obtida

\begin{equation}
{\mathcal M} = {\mathcal M_{+}} + {\mathcal M_{-}} = \frac{x^{2} + 2}{x\sqrt{x^{2}+4}}. \label{Eq24}
\end{equation}


\begin{thebibliography}{30}

%\bibitem{E1919} A. S. Eddington, ``The total eclipse of 1919 May 29 and the influence of gravitation on light", The Observatory 42, 119 (1919).

\bibitem{Apais} A. Pais, \textit{Sútil é o Senhor, Vida e Pensamento de Albert Einstein},  trad. de F. Parente e V. Esteves (Gradiva, Lisboa, 1982), p. 372-374.

\bibitem{Crispino2019} L. C. B. Crispino, D. J. Kennefick. \textit{A hundred years of the 
first experimental 
test of general relativity}, Nature Physics \textbf{15}, 416 (2019).

\bibitem{ELTurner} E. L. Turner, \textit{Gravitational lensing}, Scientific American, \textbf{259}, 26 (1988). 

%\bibitem{PLW2001} A. O. Petters, H. Levine, J. Wanbsganss, \textit{Singularity Theory and Gravitational} Lensing, Birkhauser, Berlim (2001).

\bibitem{SEF1999} P. Schneider, J. Ehlers, E. E. Falco,  \textit{Gravitational Lenses: Strong, Weak and Micro}, Saas-Fee Adv Courses vol33 (Springer,  2a. Ed., Berlim 1999).

\bibitem{Mol2002}  S. Mollerach, E. Roulet, \textit{Gravitational Lensing and Microlensing}, World Scientific, Singapore (2002).

\bibitem{CK2018} A. B. Congdon, C. R. Keeton, \textit{Principles of Gravitational Lensing} (Springer, USA 2018).

\bibitem{NB1996} R. Narayan, M. Bartelmann, \textit{Lectures on gravitational lensing},  astro-ph/9606001 (1996). 
\bibitem{Schneider2006} P. Schneider, C. Kochanek, J. Wambsganss, \textit{Gravitational Lensing: Strong, Weak and Micro} (Springer,  Berlim 2006).

\bibitem{B2010} M.  Bartelmann, \textit{Gravitational lensing}, Class. and Quant. Grav. \textbf{27}, 23 (2010).

\bibitem{T2018} Y. Tsapras, \textit{Microlensing Searches for Exoplanets}, Geosciences  \textbf{ 8(10)}, 365 (2018).

\bibitem{MN2007} H. M. Nussenzveig, \textit{Light tunneling}, em Progress in Optics 50, Ed. Emil Wolf (Elsevier 2007), p. 185

\bibitem{RS2015} R. A. Martins,  C. C. Silva, \textit{As pesquisas de Newton sobre a luz: Uma visão histórica}, Revista Brasileira de Ensino de Física, \textbf{37}, n. 4, 4202 (2015).

\bibitem{N1704} I. Newton, \textit{Optikcs or A Treatise of the Reflections, Refractions, Inflections or Colors of Light} (London 1704). A  4a. edição (1730) foi reeditada pela Dover Publications (New York, 1952), com prefácio de A. Einstein e uma introdução de E. Whitakker.  Traduzida no Brasil sob o título Ótica - por A. K. T. Assis (Edusp, São Paulo, 1996).

\bibitem{M1783} J. Michel, \textit{On the means of discovering the distance and  magnitude of the fixed stars, in consequence of the diminution of the velocity of their light...}, Philo. Trans. R. Soc. Lon. \textbf{74}, 35  (1784). 
%Reimpresso por C. Misner, K. Thorne e J. Wheeler, \textit{Black Holes: Selected Reprints} (Ed. S. Detweiler, Am. Assoc. of Physics Teachers, Stony Brook 1982). 

\bibitem{L1798} P. S. Laplace, \textit{Exposition du Systéme du Monde}, vol. \textbf{II},  (De I'Imprimerie du Circle Social, Paris, 1796), p. 305.

\bibitem{S1801} J. Soldner, \textit{On the deflection of a ray of light from its straight-line motion through the attraction of a world-body on which it passes close by},  https://en.wikisource.org/wiki/Translation: On the Deflection of a Light Ray from its Rectilinear Motion -
Traduzido do alemão, Berl. Astron. Jahrb.  \textbf{161} (1804).

\bibitem{LL1975} L. D. Landau, E. M. Lifshitz, \textit{The Classical Theory of Fields} (Pergamon  Press, USA, 1975), p. 301.

\bibitem{W1986} C. M. Will,  \textit{Henry Cavendish, Johann von Soldner, and the deflection of light},  Amer. J.  Phys. \textbf{56}, 413 (1988).

\bibitem{E1905} A. Einstein,  \textit{Sobre a eletrodinâmica dos corpos em movimento}, em \textit{Textos Fundamentais da Física Moderna}, (Fundação Calouste Gulbenkian, Lisboa 1971), p. 47. Trad. do original alemão, Ann. der phys. \textbf{17} (1905). 

\bibitem{P1905} H. Poincaré, \textit{Sur la dinamique du Eletron}, C. R. Acad. Sci. Paris  \textbf{14}, 1504 (1905); Oeuvers Completes, (Gauthier Villars, Paris \textbf{9}, 489, 1954).

\bibitem{E1907} A. Einstein, Jahrb. Radioakt. ElektroniK \textbf{4}, 411 (1907). Correções de Einstein em Jahrb. Radioakt. ElektroniK \textbf{5}, 512 (1908). A parte relacionada com a gravitação pode  ser lida (com notas) em  H.M. Schwartz, Amer. J. Phys. \textbf{45}, 899 (1977). 

\bibitem{E1911} A. Einstein, \textit{Sobre a Influência da Gravidade na Propagação da Luz}, coleção \textit{Textos Fundamentais de Física Moderna}, Volume \textbf{I}, O Princípio da Relatividade (Fundação Calouste Gulbenkian, 1971), p. 99.  Traduzido do  original alemão: Annalen der Physik \textbf{35}, 898 (1911).

\bibitem{Dys20} F. W. Dyson,  A. S. Eddington,  C. R. Davidson, \textit{A determination of the deflection of light by the sun's gravitational field from observations made at the total eclipse of May 29, 1919}, Mem. Roy. Astron. Soc. \textbf{ 62}, 291  (1920).

\bibitem{Leb95} D. E. Lebach  \textit{et al.}, \textit{Measurement of the solar gravitational deflection of radio waves using very-long-baseline	interferometry}, Phys. Rev. D \textbf{75}, 1439 (1995).	

\bibitem{C1924} O. Chwolson, \textit{Über eine mögliche form fiktiver Doppelsterne},  Astr. Nachrichten \textbf{221}, 329 (1924).

\bibitem{E1936} A. Einstein , \textit{Lens-like action of a star by deviation of light in the gravitational field}, Science \textbf{84}, 506 (1936).

\bibitem{Renn1997} J. Renn, T. Sauer, J. Stachel,  \textit{The origin of gravitational lensing: A  postscritpt to Einstein's 1936},  Science \textbf{ 275}, 184 (1997).


\bibitem{Z1937a} F. Zwicky, \textit{Nebulae as gravitational lenses}, Phys. Rev. \textbf{51}, 290 (1937).

\bibitem{Z1937b}  F. Zwicky , \textit{On the probability of detecting nebulae which act as gravitational lenses}, Phys. Rev. \textbf{51}, 679 (1937).

\bibitem{LA2000} J. A. S. Lima, J. S. Alcaniz, \textit{Angular size in quintessence cosmology},  Astron. Astrophys. \textbf{357}, 393  (2000), astro-ph/0003189

\bibitem{Santos2007} R. C. Santos, \textit{Efeitos das inomogeneidades da matéria em Cosmologias Aceleradas}, Tese de Doutorado (2007). 

\bibitem{SL2008} R. C. Santos,  J. A. S. Lima, \textit{Clustering, angular size and dark energy},  Phys. Rev. D \textbf{77}, 083505  (2008), arXiv:0803.1865 

\bibitem{LS2018} J. A. S. Lima,  R. C. Santos, \textit{100 Anos da cosmologia relativística (1917-2017). Parte I: Das origens à descoberta da expansão universal (1929)}, Revista Brasileira de Ensino de Física \textbf{40}, 1, e1313 (2018), arXiv:1709.03693 [astro-ph.CO].

\bibitem{LNT2017} H. Ebiling \textit{et al.},  \textit{Thirty-fold: Extreme gravitational lensing of a quiescent Galaxy at $z=1.6$ }, Astrophys. J. \textbf{852}, L7 (2017).


\bibitem{SEF1999c} P. Schneider, J. Ehlers, E. E. Falco, op. cit. \cite{SEF1999}, p. 29-31

\bibitem{Mol2002b}  S. Mollerach, E. Roulet, op. cit. \cite{Mol2002}, p. 33.

\bibitem{B1981} W. L. Burke, \textit{Multiple gravitational imaging by distributed masses}, Astrophys. J. Lett. \textbf{244}, 1 (1981). 

\bibitem{M1985} R. H. McKenzie, \textit{A gravitational lens produces an odd number of images}, JMP \textbf{26}, 1592 (1985).

\bibitem{LP1986} R. Lynds, V. Petrosian, \textit{Giant luminous arcs in galaxy clusters}, BAAS \textbf{18}, 1014 (1986).

\bibitem{T1990} J. A. Tyson, F. Valdes, R. A. Wenk, \textit{Detection of systematic gravitational lens galaxy image alignments - Mapping dark matter in galaxy clusters}, Astrophys. J.  \textbf{349}, L1 (1990). 

\bibitem{SEF1999b} P. Schneider, J. Ehlers, E. E. Falco, op. cit. \cite{SEF1999}, p. 43-44.

\bibitem{Schmidt1963}  M. Schmidt,  \textit{3C 273: a star-like object with large redshift}, Nature \textbf{197} (4872): 1040 (1963). 

\bibitem{Re1964a}  S. Refsdal, \textit{The gravitational lens effect}, MNRAS \textbf{128}, 295 (1964).

\bibitem{Re1964b} S. Refsdal, \textit{On the possibility of determining Hubble's parameter and the masses of galaxies from the gravitational lens effect}, MNRAS \textbf{128}, 307 (1964).

\bibitem{WCW1979} D. Walsh, R. F. Carswell, R. J. Weymann, \textit{0957 + 561 A, B: Twin quasistellar objects or gravitational lensing?} Nature \textbf{279}, 381 (1979).

\bibitem{Hewitt1988} J. Hewitt, \textit{The unusual radio source MG $1131+0456$: A possible Einstein ring}, Nature \textbf{ 333}, 537 (1988).

\bibitem{Y1980}  P. Young, \textit{et al.}., \textit{The double quasar Q0957 + 561 A, B - A gravitational lens image formed by a galaxy at Z = 0.39},  Astrophys. J. \textbf{241}, 507 (1980).

\bibitem {KF70}K. C. Freeman, \textit{On the disks of spiral and S0 Galaxies}, ApJ \textbf{160}, 811; K. C. Freeman, \textit{Erratum: on the disks of spiral and s0 Galaxies},  Astrophys. J. \textbf{161}, 802 (1970).

\bibitem{VR} V. C. Rubin, W. K. Ford Jr. , \textit{Rotation of the Andromeda nebula from a spectroscopic survey of emission regions},  Astrophys. J. \textbf{159}, 379 (1970); V. C. Rubin, \textit{The rotation of spiral Galaxies}, Science  \textbf{220}, 1339 (1983).

\bibitem{VR2} V. C. Rubin, \textit{Dark matter in spiral Galaxies}, Scientific American \textbf{248}, 88 (1983). 

\bibitem{OP73} J. P. Ostriker, P. J. E. Peebles,  \textit{A numerical study of the stability of flattened Galaxies: or, can cold Galaxies survive?}, Astrophys. J. \textbf{ 186}, 467 (1973).

\bibitem{P1993} P. J. E. Peebles,  \textit{Principles of Physical Cosmology}, Princeton University Press, New Jersey, p. 47 (1993).

\bibitem{DM2019} B. K. Stoychev, \textit{Clues to the nature of dark matter from first Galaxies}, arXiv:1905.00432 [astro-ph.GA] (2019).

\bibitem{W2006b} J. Wambsganess, \textit{Gravitational Microlensing}, op. cit. \cite{Schneider2006}, p. 453.

\bibitem{BP1986a} B. Paczynski, \textit{Gravitational microlensing at large optical depth}, Astrophys. J.  \textbf{301}, 503 (1986).

\bibitem{BP1986b} B. Paczynski,  \textit{Gravitational microlensing by the galactic halo}, Astrophys. J.  \textbf{304}, 1 (1986).

\bibitem{BP1996} B. Paczynski, \textit{Gravitational microlensing in the local group}, Ann. Rev. Astron. Astrophys. \textbf{34}, 419 (1996).


\bibitem{Beaulieu2006} J. P. Beaulieu  \textit{et al.},  \textit{Discovery of a cool planet of 5.5 Earth masses through gravitational microlensing},  Nature \textbf{439}, 437 (2006). 

\bibitem{S2003} K. C. Sahu, \textit{Microlensing towards the Magellanic Clouds: Nature of the Lenses and Implications for Dark Matter}, Proceedings of the STScI Symposium on "Dark Universe: Matter, Energy, and Gravity", M. Livio (ed.), Cambridge Univ. Press: Cambridge, p. 14 (2003).

\bibitem{MP1991} S. Mao and B. Paczynski, \textit{Gravitational Microlensing by double Stars and Planetary systems}, Astrophys. J. 374, L37 (1991).

\bibitem{B2004} I. A. Bond \textit{et al.},  \textit{OGLE 2003-BLG-235/MOA 2003-BLG-53: A Planetary Microlensing event}, Astrophys. J. \textbf{606}, L155 (2004).

\bibitem{Oliveira2017}  R. Monteiro-Oliveira, \textit{et al.}, \textit{The merger history of the complex cluster Abell 1758: a combined weak lensing and spectroscopic view}, MNRAS \textbf{466}, 2614 (2017).

\bibitem{CL2004} E. S. Cypriano \textit{et al.} \textit{Weak-lensing mass distributions for 24 X-ray Abell clusters}, Astrophys. J. \textbf{613}, 95

%J. P. Kneib, L. E. Campusano

\bibitem{Meneghetti2013} M.  Meneghetti \textit{et al.}, \textit{Arc statistics},  Space Sci. Rev. \textbf{177}, 31 (2013).

\bibitem{Dodelson2017} S. Dodelson, \textit{Gravitational Lensing} (Cambridge University Press 2017).

\bibitem{Z33} F. Zwicky, \textit{Die Rotverschieb ung von extragalaktischen Nebeln}, Helvetica Phys. Acta \textbf{6}, 110 (1933). 

\bibitem{Clowe} D. Clowe, \textit{et al.}, \textit{A direct empirical proof of the existence of dark matter}. Astrophys. J. Lett., \textbf{648}, L109 (2006).

\bibitem{Oliveira2018} R. Monteiro-Oliveira, \textit{et al.}, \textit{New insights on the dissociative merging galaxy cluster Abell 2034}. MNRAS \textbf{481},  1097 (2018).

\bibitem{Markevitch} M. Markevitch \textit{et al.},  \textit{Direct constraints on the dark matter self-interaction cross-section from the merging galaxy cluster 1E0657-56}, Multiwavelength Cosmology, editado por Manolis Plionis, Kluwer Academic Publishers, USA, p. 263 (2004).

\bibitem{M1983} M. Milgrom, \textit{A modification of Newtonian dynamics as a possible alternative to the hidden matter hypothesis}, Astrophys. J. \textbf{270}, 365 (1983). 

\bibitem{MOND2001} A. Aguirre, J. Schaye, E. Quataert,  \textit{Problems for MOND in Clusters and the Ly-alpha Forest}, Astrophys. J. \textbf{561}, 550 (2001)


\bibitem{Riess1998} A. G. Riess \textit{et al.}, \textit{Observational evidence from supernovae for an accelerating Universe and a
cosmological constant}, Astron. J.
 \textbf{ 116}, 1009, (1998).

\bibitem{Perlmutter1999} S. Perlmutter \textit{et al.}, \textit{Measurements of $\Omega$ and $\Lambda$ from 42 high-redshift Supernovae}, Astrophys. J. \textbf{517}, 565 (1999).

\bibitem{DETF2006} A. Albrecht et al., \textit{Report on the dark Energy Task Force}, arXiv:astro-ph/0609591, DOI:$  10.2172/897600$ (2006).

\bibitem{Wang05} Y. Wang, \textit{Observational signatures of the weak lensing magnification of
  Supernovae},  JCAP \textbf{2005}, 005 (2005).
  
  \bibitem{R2017} Rachel Mandelbaum, \textit{Weak lensing for precision
cosmology}, ARAA  \textbf{56}, 393 (2017).

%\bibitem{Metcalf99} R. Benton, \textit{Gravitational lensing of high-redshift Type IA supernovae: a probe
% of medium-scale structure},%MNRAS\textbf{305}, 746 (1999).

\bibitem{Goliath00} M. Goliath, E. M{\"o}rtsell, \textit{Gravitational lensing of Type Ia Supernovae}, Phys. Lett. B \textbf{486}, 249 (2000).

\bibitem{Jonsson08} J. J{\"o}nsson \textit{et al.}, \textit{Prospects and pitfalls of gravitational lensing in large Supernova   surveys}, Astron. \& Astrophys.  \textbf{487}, 467 (2008).

\bibitem{Smith14} M. Smith \textit{et al.}, \textit{The effect of weak lensing on distance estimates from Supernovae},  Astrophys. J. \textbf{780}, 24  (2014).

\bibitem{Kelly2015} P. I. Kelly  \textit{et al.}, \textit{Multiple images of a highly magnified Supernova formed by an early-type Cluster Galaxy lens}, Science \textbf{347}, 1123 (2015).

\bibitem{Treu2015} T. Treu, R. S. Ellis, \textit{Gravitational lensing - Einstein's unfinished symphony}, Contemp. Phys. \textbf{56}, 17 (2015).

\bibitem{Treu2016} T. Treu \textit{et al.}, \textit{"Refsdal" Meets Popper: Comparing predictions of the re-appearance of the multiply imaged Supernova behind MACSJ1149.5+2223},  Astrophys. J. \textbf{817}, 60 (2016). 

\bibitem{Grillo2018} C.  Grillo \textit{et al.}, \textit{Measuring the value of the Hubble constant "à la Refsdal"},   \textbf{860}, 94 (2018).

\bibitem{Riess2011} A. G. Riess \textit{et al.},  \textit{A 3\% Solution: Determination of the Hubble Constant with the Hubble Space Telescope and Wide Field Camera 3}, Astrophys. J \textbf{730} 119  (2011).
Erratum: Astrophys. J \textbf{732},  129 (2011).
 
\bibitem{LC2014} J. A. S Lima, J. V. Cunha,  \textit{A 3\% Determination of H0 at Intermediate Redshifts}, Astrophys. J. \textbf{ 781}, 2, L38  (2014), arXiv:1206.0332 [astro-ph.CO]


\bibitem{Freedman2017} W. L. Freedman, \textit{Cosmology at a Crossroads}, Nat. Astron. \textbf{1},  0121 (2017).

\bibitem{planck2015} Planck Collaboration, \textit{ XIII. Cosmological parameters}, Astron. \& Astrophys. \textbf{ 594}, A13 (2016).

\bibitem{Riess2016}  A. G. Riess \textit{et al.},  \textit{ A 2.4\% determination of the local value of the Hubble constant.} Astrophys. J.  \textbf{ 826}, 56 (2016).

\bibitem{planck2018} Planck Collaboration, \textit{Planck 2018 results. VI. Cosmological parameters}, arXiv:1807.06209 (2018).
%\bibitem{Riess2018} A. G. Riess \textit{et al.},  \textit{ Milky Way Cepheid Standards for Measuring Cosmic Distances and Application to Gaia DR2: Implications for the Hubble Constant}, Astrophys. J. \textbf{861}, 126 (2018).

\bibitem{riess2019}  A. G. Riess \textit{et al.}, \textit{Large Magellanic Cloud Cepheid standards provide a 1\% foundation for the determination of the Hubble constant and stronger evidence for physics beyond $\Lambda$CDM}, Astrophys. J. \textbf{876}, 1, 85  (2019).  

\bibitem{Suyu2017} S. H. Suyu  \textit{et al.},  \textit{H0LiCOW - I. H0 lenses in COSMOGRAIL's Wellspring: program overview}, MNRAS \textbf{468}, 2590 (2017).

\bibitem{Courbin2018}  F. Courbin \textit{et al.}, \textit{ COSMOGRAIL: the COSmological MOnitoring of GRAvItational Lenses XVI. Time delays for the quadruply imaged quasar DES J0408 5354 with high cadence photometric monitoring}, Astron.  Astrophys. \textbf{609}, A71 (2018). 

\bibitem{Birrer2019}  S. Birrer \textit{et al.}, \textit{H0LiCOW - IX. Cosmographic analysis of the doubly imaged quasar SDSS 1206+4332 and a new measurement of the Hubble constant},  MNRAS \textbf{484}, 4726 (2019).

\bibitem{Abbott2016a} B. P. Abbott {\textit{et al.}} {\textit{Observation of gravitational waves from a binary Black Hole Merger}} (LIGO Scientific Collaboration and Virgo Collaboration), Phys. Rev. Lett.  \textbf{116}, 061102  (2016).

\bibitem{Abbott2016b} B. P. Abbott  {\textit{et al.}}, \textit{GW151226: Observation of gravitational waves from a 22-Solar-Mass binary Black Hole coalescence}(LIGO Scientific Collaboration and Virgo Collaboration), Phys. Rev. Lett. \textbf{116}, 241103 (2016).

\bibitem{Abbott2017}
 B. P. Abbott {\textit{et al.}} \textit{GW170104: Observation of a 50-Solar-Mass binary Black Hole coalescence at redshift 0.2} (LIGO Scientific Collaboration and Virgo Collaboration),  Phys. Rev. Lett. \textbf{118}, 221101 (2017). 


\bibitem{L2017} Liao K. \textit{et al.}, \textit{Precision cosmology from future lensed gravitational wave and electromagnetic signals}, Nature Communications \textbf{8}, 1148 (2017).

\bibitem{F2019} S. M.  Feeney \textit{et al.}, \textit{Prospects for Resolving the Hubble constant tension with standard sirens},   Phys.  Rev.  Let.  \textbf{122}, 061105 (2019).

\bibitem{CSL2018} J. V. Cunha,  F. E. Silva, J. A. S. Lima  \textit{Gravitational waves from ultra-short period Exoplanets}, MNRAS \textbf{480}, L28 (2018), arXiv:1807.04877 [astro-ph.EP]

\bibitem{Tamanini2019} N. Tamanini, C. Danielski,  \textit{The gravitational wave detection of exoplanets orbiting white dwarf binaries using LISA}, Nature Astronomy (2019), online, July 10.


\end{thebibliography}
\end{document}